\documentclass{aa}  

\usepackage{graphicx}
\usepackage{txfonts}
\usepackage[T1]{fontenc}
\usepackage{ae,aecompl}

\usepackage{tablefootnote}

\usepackage{amsmath}
\usepackage{amssymb}
\usepackage{multirow,bigdelim}
\usepackage{mathtools}
\usepackage{siunitx}
\usepackage[colorlinks=true,linkcolor=blue, citecolor=blue]{hyperref}%

\usepackage{subfig}

\usepackage{ulem}

\usepackage{ulem}

\newcommand{\wgx}{w_{\rm{g}\times}}
\newcommand{\wgp}{w_{\rm{g+}}}
\newcommand{\xigp}{$\xi_{\rm{g+}}$ }
\newcommand{\xigg}{$\xi_{\rm{gg}}$ }
\newcommand{\mpch}{\,h^{-1}{\rm{Mpc}}}
\newcommand{\dd}{\mathrm{d}}

\newcommand{\lum}{\texttt{luminous}}
\newcommand{\dense}{\texttt{dense}}
\newcommand{\lensfit}{\textit{lens}fit}

\usepackage{bm}
\usepackage{soul}
\usepackage{comment}

\usepackage{color}

\numberwithin{equation}{section}

\usepackage{etoolbox}

\begin{document}

\title{KiDS-1000: Constraints on the intrinsic alignment of luminous red galaxies}
\titlerunning{KiDS-1000 LRG intrinsic alignments}

\author{
Maria Cristina Fortuna\inst{1}\thanks{\emph{E-mail:} fortuna@strw.leidenuniv.nl},
Henk Hoekstra\inst{1},
Harry Johnston\inst{2},
Mohammadjavad Vakili\inst{1},
Arun Kannawadi\inst{3},
Christos Georgiou\inst{1},
Benjamin Joachimi\inst{4},
Angus H. Wright\inst{5},
Marika Asgari\inst{6},
Maciej Bilicki\inst{7},
Catherine Heymans\inst{5,6},
Hendrik Hildebrandt\inst{5},
Konrad Kuijken\inst{1},
Maximilian Von Wietersheim-Kramsta\inst{4}
  }

\authorrunning{M.C. Fortuna et al. }

\institute{Leiden Observatory, Leiden University, PO Box 9513, Leiden, NL-2300 RA, the Netherlands \and
Institute for Theoretical Physics, Utrecht University, Princetonplein 5, 3584 CE Utrecht, The Netherlands \and
Department of Astrophysical Sciences, Princeton University, 4 Ivy Lane, Princeton, NJ 08544, USA \and
Department of Physics and Astronomy, University College London, Gower Street, London WC1E 6BT, UK \and
Ruhr University Bochum, Faculty of Physics and Astronomy, Astronomical Institute (AIRUB), German Centre for Cosmological Lensing, 44780 Bochum, Germany \and
Institute for Astronomy, University of Edinburgh, Royal Observatory, Blackford Hill, Edinburgh, EH9 3HJ, UK \and
Center for Theoretical Physics, Polish Academy of Sciences, al. Lotników 32/46, 02-668 Warsaw, Poland
}

\date{Accepted XXX. Received YYY; in original form ZZZ}

\label{firstpage}
\makeatletter
\renewcommand*\aa@pageof{, page \thepage{} of \pageref*{LastPage}}
\makeatother

\abstract{
We constrain the luminosity and redshift dependence of the intrinsic alignment (IA) of a nearly volume-limited sample of luminous red galaxies selected from the fourth public data release of the Kilo-Degree Survey (KiDS-1000).
To measure the shapes of the galaxies, we used two complementary algorithms, finding consistent IA measurements for the overlapping galaxy sample. The global significance of IA detection across our two independent luminous red galaxy samples, with our favoured method of shape estimation, is $\sim10.7\sigma$. We find no significant dependence with redshift of the IA signal in the range $0.2<z<0.8$, nor a dependence with luminosity below $L_r\lesssim 2.9 \times 10^{10} h^{-2} L_{r,\odot}$. Above this luminosity, however, we find that the IA signal increases as a power law, although our results are also compatible with linear growth within the current uncertainties. This behaviour motivates the use of a broken power law model when accounting for the luminosity dependence of IA contamination in cosmic shear studies.}

\maketitle



\section{Introduction}

Galaxies that form close to a matter over-density are affected by the tide induced by the quadrupole of the surrounding gravitational field, and the distribution of stars will adjust accordingly. This process, which starts during the initial stages of galaxy formation \citep{Catelan2001}, can persist over their entire lifetime, as galaxies have continuous gravitational interactions with the surrounding matter \citep[e.g.][]{Bhowmick2019}, and leads to the intrinsic alignment (IA) of galaxies.

This tendency of neighbouring galaxy pairs to have a similar orientation of their intrinsic shapes is an important contaminant for weak gravitational lensing measurements \citep[e.g.][]{Joachimi2015review}. The matter distribution along the line-of-sight distorts the images of background galaxies, resulting in apparent correlations in their shapes. Intrinsic alignment contributes to the observed correlations, complicating the interpretation. To infer unbiased cosmological parameter estimates it is therefore crucial to account for the IA contribution. This is particularly important in the light of future surveys, such as \textit{Euclid}\footnote{\url{https://www.euclid-ec.org}} \citep{Laureijs2011} and the Large Synoptic Survey Telescope (LSST)\footnote{\url{https://www.lsst.org}} at the Vera C. Rubin Observatory \citep{Abell2009lsst}, which aim to constrain the cosmological parameters with sub-percent accuracy \citep[for a forecast of the IA impact on current and upcoming surveys see][among others]{Kirk2010, Krause2016}. Some recent results on current weak lensing studies are available in, for example, \citet[][]{Aihara2018HSC,Asgari2021, Abbott2021DESY3}.

To provide informative priors to lensing studies, it is essential to learn as much as possible from direct observations of IA. It is, however, also important that such results can be related to the properties of galaxies that give rise to the alignment signal in cosmic shear surveys \citep{fortuna2020halo}. Intrinsic alignment studies are typically limited to relatively bright galaxies, which often sit at the centre of their own group or cluster, and it is thus possible to connect their alignment to the underlying dark matter halo alignment via analytic models \citep{Hirata2004}. The picture becomes more complicated when considering samples that contain a significant fraction of satellite galaxies: The alignment of satellites arises as a result of the continuous torque exercised by the intra-halo tidal fields while the satellite orbits inside the halo \citep{Pereira2008, Pereira2010}. This leads to a radial alignment, which also depends on the galaxy distance from the centre of the halo \citep{Georgiou2019b}. At the same time, satellites fall into halos through the filaments of the large-scale structure, and this persists as an anisotropic distribution within the halo, which has been detected both in simulations \citep{Knebe2004, Zentner2005} and observations \citep{West2000VirgoCluster, Bailin2008, Huang2016, Johnston2019, Georgiou2019b}. The combination of these two effects complicates the picture. At small scales, where the satellite contribution is expected to be important, their signal may be described using a halo model formalism \citep{SchneiderBridle2010, fortuna2020halo}, but their contribution to IA on large scales remains poorly constrained \citep{Johnston2019}; although it is expected that they are not aligned, they do affect the inferred amplitude because they contribute to the overall mix of galaxies. This prevents a straightforward interpretation of any secondary sample dependence of the IA signal sourced by the central galaxy population, such as the dependence on luminosity or colour, in mixed samples where the fraction of satellites is relevant.

Observational studies have found discordant results regarding the presence of a luminosity dependence of the IA signal, with the bright end being well described by a steep power law with index $\sim 1.2$ \citep{Hirata2007, Joachimi2011b, Singh2015}, while less luminous galaxies do not show any significant dependence of the IA signal with luminosity \citep{Johnston2019}. A recent investigation using hydrodynamic simulations by \cite{Samuroff2020} supports a flatter slope, in agreement with \citet{Johnston2019} and \citet[][]{fortuna2020halo} at low luminosities but in tension with previous studies that probe more luminous galaxies. The interpretation of these results is also affected by the presence of satellites, whose fraction varies with luminosity and depends on the specific selection function of the data. At low redshift, a cosmic shear survey is dominated by faint galaxies, and improving our understanding of the  IA signal at low luminosities is one of the most urgent questions for IA studies.

Another relevant aspect that is often neglected is the dependence of IA on the shape measurement method \citep{Singh2016}. The tendency to align in the direction of the surrounding tidal field is a function of galaxy scale \citep{Georgiou2019b}, with the outermost parts -- which are more weakly gravitationally locked to the galaxy -- showing a more severe twist. It increases the IA signal associated with shapes measured via algorithms that assign more importance to the galaxy outskirts. In contrast, lensing studies typically prefer shape methods that give more weight to the inner part of a galaxy. Accounting for this discrepancy is potentially relevant for future cosmic shear studies.

In this work we focus on investigating the luminosity dependence of the IA signal in the least constrained regime, $M_r\gtrsim-22$. We employ two different samples, which differ in mean luminosity and number density. We limit the analysis to the large-scale alignment, for which a theoretical framework is already available and where the luminosity dependence is known to play a crucial role \citep{fortuna2020halo}. We also provide estimates of the satellite fractions present in our samples in order to guide future work on the modelling of satellite alignment at large scales.
We also explore the dependence of our signal on the shape measurement algorithm used to create the shape catalogue. We compare the signal as measured by two complementary algorithms: \textsc{DEIMOS} \citep[DEconvolution In MOment Space; ][]{Melchior2011}, which has been widely used in IA studies \citep{Georgiou2019, Johnston2019, Georgiou2019b}, and \lensfit\ \citep{Miller2007, Miller2013} which has been used for the cosmological analysis of the Canada-France-Hawaii Telescope Lensing Survey \citep[CFHTLenS;][]{Heymans2013CFHTLens} and the Kilo-Degree Survey \citep[KiDS; see][and references therein]{Asgari2021}.

One of the main limitations for measuring IA is the necessity of simultaneously relying on high-quality images and precise redshifts to properly identify physically close pairs of galaxies that share the same gravitational tidal shear. Wide field image surveys provide high-quality images, but the uncertainty in the photometric redshifts is too large for
useful IA measurements. Fortunately, using a specific selection in colours, it is possible to obtain a sub-sample of galaxies with more precise photometric redshifts: the luminous red galaxies (LRGs). At any given redshift, LRGs populate a well-defined region in the colour-magnitude diagram, known as the red-sequence ridgeline. Using this unique property, it is possible to design a specific algorithm to select LRGs in photometric surveys, which results in both precise and accurate redshifts \citep{Rozo2016, Vakili2019, Vakili2020}. Luminous red galaxies have also been shown to be strongly affected by the surrounding tidal fields, making them an extremely suitable sample for exploring the behaviour of IA at different redshifts and as a function of secondary galaxy properties, such as luminosity and type (central or satellites).

\citet{Joachimi2011b} first studied the IA signal of an LRG sample with photometric redshifts. In this paper we follow their main approach but use a catalogue of LRGs selected by \citet[]{Vakili2020} using the KiDS fourth public data release \citep[KiDS-1000][]{Kuijken2019DR4}.

The paper is structured as follows. In Sect. \ref{sec:kids} we describe our data and the characteristics of our two main samples. In Sect.~\ref{sec:shape_measurements} we introduce the two shape measurement methods employed in the analysis and present the strategy adopted to calibrate the bias in the measured shapes. Section ~\ref{sec:correlation_function_measurements} presents the estimators we use to extract the signal from the data, while Sect.~\ref{sec:theoretical_framework} illustrates the theoretical framework we rely on when modelling the signal: the way the model accounts for the use of photometric redshifts as well as the way we account for astrophysical contaminants. Finally, we present our main results in Sect.~\ref{sec:results} and conclude in Sect.~\ref{sec:conclusions}.

Throughout the paper, we assume a flat $\Lambda$ cold dark matter cosmology with $h=0.7, \Omega_{\rm m} = 0.25, \Omega_{\rm b} = 0.044, \sigma_8 = 0.8$, and $n_{\rm s} = 0.96$. 

\section{KiDS}
\label{sec:kids}

The Kilo-Degree Survey is a multi-band imaging survey designed for weak lensing studies, currently at its fourth data release \citep[KiDS-1000;][]{Kuijken2019DR4}. The data are obtained with the OmegaCAM instrument \citep{Kuijken2011} on the VLT Survey Telescope \citep[VST;][]{Capaccioli2012}. This combination of telescope and camera was designed specifically to produce high-quality images in the $ugri$ filters, with best seeing-conditions in the $r-$band, and a mean magnitude limit of $\sim 25$ ($5\sigma$ in a $2''$ aperture). These measurements are combined with results from the VISTA Kilo-degree INfrared Galaxy survey
\citep[VIKING;][]{Edge2013}, which surveyed the same area in five infrared bands ($ZYJHK_\mathrm{s}$). This resulted in high-quality photometry in nine bands across approximately $1000 {\rm \ deg}^2$ imaged by the fourth data release\footnote{The survey was recently completed, imaging a final total of 1350 deg$^2$.}. The VIKING data are important for the LRG selection at high redshift \citep{Vakili2020}: the $Z$ band is included in the red-sequence template and improves the constraints on the redshift of the high-redshift galaxies, while the $K_\mathrm{s}$ band allows for a clean separation between galaxies and stars in the $(r-K_\mathrm{s})-(r-z)$ colour-colour space.

\subsection{The LRG sample}
\label{sec:lrg_sample}

Red-sequence galaxies are characterised by a tight colour-redshift relation, so that at any given redshift they follow a narrow ridgeline in the colour-magnitude space. This relation can be exploited to select red galaxies from photometric data and obtain precise photometric redshifts. Here we use the catalogue of LRGs presented in \cite{Vakili2020}. It uses a variation of the \textsc{redMagiC} algorithm \citep{Rykoff2014} to select LRGs from the KiDS-1000 data. As detailed in \citet[][]{Vakili2019} and \citet[][]{Vakili2020}, the red-sequence template is calibrated using the regions of KiDS that overlap with a number of spectroscopic surveys: SDSS DR13 \citep{Albareti2017}, 2dFLenS \citep{Blake2016}, GAMA \citep{Driver2011}, together with the GAMA G10 region, which overlaps with COSMOS \citep{Davies2015}.

\begin{figure}
\centering
\includegraphics[width=\columnwidth]{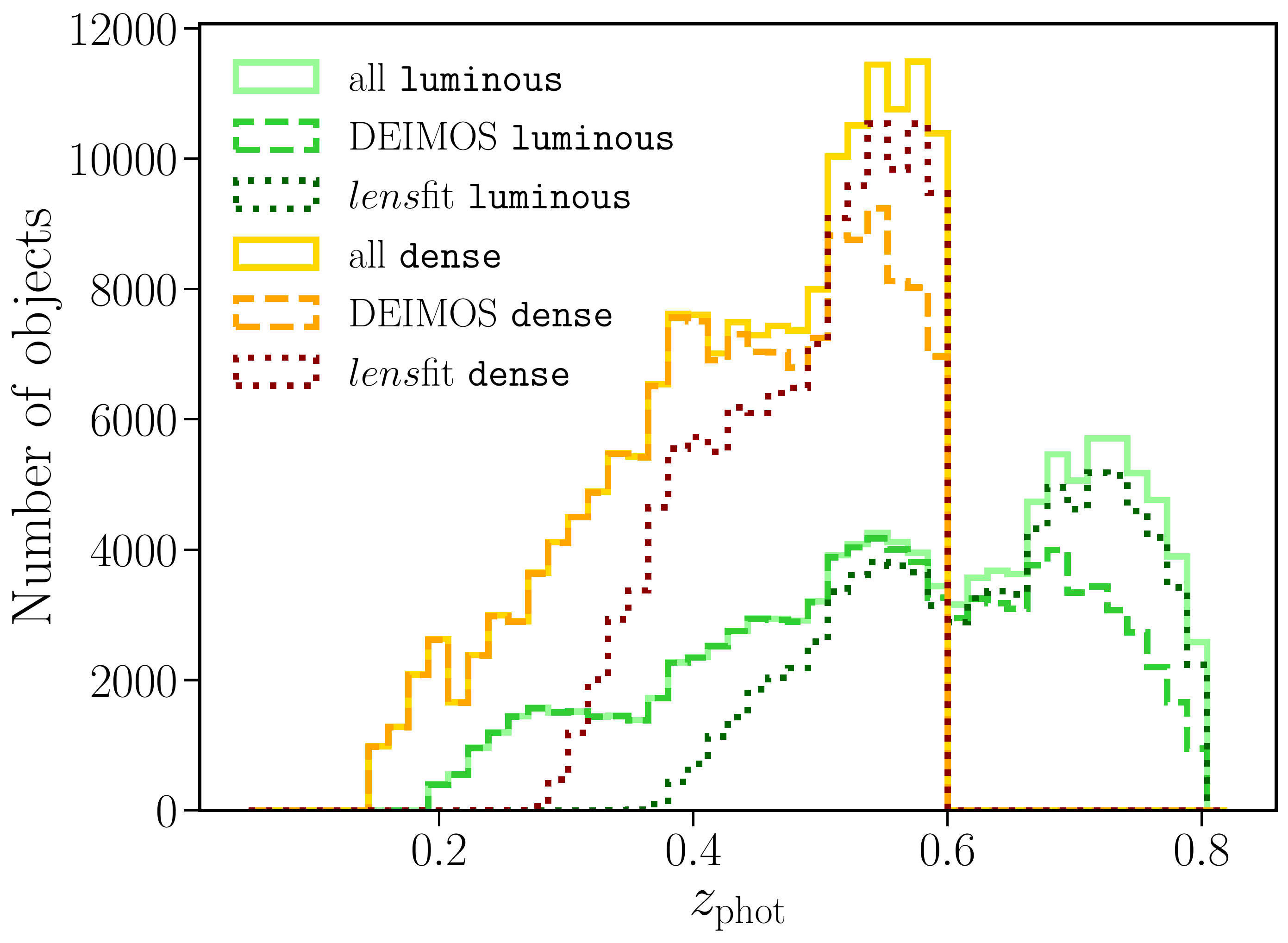}
\caption{Photometric redshift distributions for our density (all) and shape catalogues (\lensfit\ and \textsc{DEIMOS}; see text for details). The orange histograms show the distribution for the \dense\ samples, which is limited to $z_{\rm phot}<0.6$, whereas the \lum\ sample (green) is restricted to $z_{\rm phot}<0.8$.}
\label{fig:z_histo_shapes}
\end{figure}

The algorithm is designed to return a sample of LRGs with a constant comoving number density. It achieves this by imposing a redshift-dependent magnitude cut that depends on $m_{r}^\mathrm{pivot}(z)$, the characteristic $r$-band magnitude of the \cite{Schechter1976} function, assuming a faint-end slope $\alpha=1$ \citep[for more details, see][sect. 3.1]{Vakili2019}. We use this to define two samples that differ from each other in terms of their minimum luminosity relative to the luminosity $L_\mathrm{pivot}(z)$. We refer to them as our \lum\ sample (high luminosity, low number density, $L_{\rm min}/L_\mathrm{pivot}(z)=1$) and \dense\ sample (lower luminosity, higher number density, $L_{\rm min}/L_\mathrm{pivot}(z)=0.5$). To ensure that the two samples are separate, we remove the galaxies in the \dense\ sample that also belong to the \lum\ one. However, this does not mean they do not overlap in their 
physical properties. In particular, they overlap partially in luminosity, a feature that we will exploit later in the paper.

\begin{figure*}
\centering
\includegraphics[width=\textwidth]{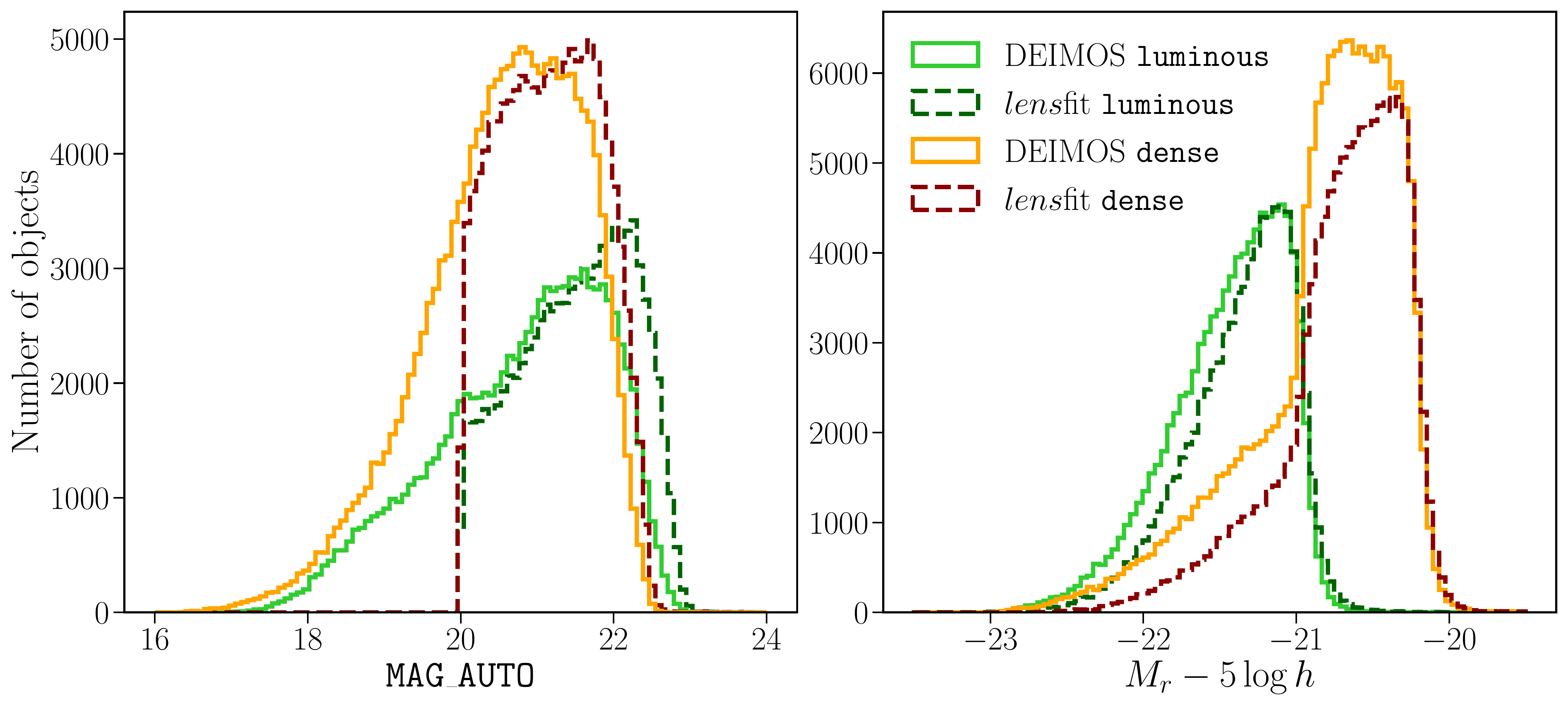}
\caption{The magnitude distributions of the samples used in the analysis. {\it Left panel:} Histograms of the apparent magnitude, {\texttt{MAG\_AUTO}} in the $r$-band for the galaxies in the \dense\ (orange lines) and \lum\ (green lines) samples with shapes measured by \lensfit\ (darker colours) and \textsc{DEIMOS} (lighter colours). {\it Right panel:} Histograms of the absolute magnitudes in the $r$-band ($K+e$ corrected) for the same samples.} \label{fig:mag_histo_dense_and_lum}
\end{figure*}

As shown in Fig.~\ref{fig:z_histo_shapes}, the two samples also span different redshift ranges. The \lum\ sample extends from $z=0.2$ to $z=0.8$. After applying a conservative mask to select only objects with a high probability to be red-sequence galaxies (corresponding to objects with a clear separation from the star sequence in the colour-colour diagram), we are left with $117\,001$ galaxies, which comprise our density sample. By density sample---not to be confused with the \dense\ sample described above---we refer to the sample used to trace galaxy positions, as opposed to the shape sample, which is the sample used for the measurement of galaxy orientations and is composed by the galaxies of the corresponding density sample for which a given shape measurement algorithm is able to measure the galaxy shape. The density and shape samples used in this analysis are visible in Fig.~\ref{fig:z_histo_shapes}, where the density samples of the \lum\ and \dense\ samples are referred to as `all' galaxies.
The \dense\ sample is obtained with the same strategy, but we further impose $z<0.6$ to ensure the completeness and purity of the sample (see Fig. 4 in \citet[][]{Vakili2020}). This leads to a final sample of $173\,445$ galaxies.
As shown in \cite{Vakili2020}, the redshift errors are well described by a Student's $t-$distribution. The width of the distribution increases slightly with redshift, with typical values around $\sigma_z \sim 0.014- 0.019$. For further details on the sample selection and redshift estimation, we refer the interested reader to \citet{Vakili2020}. 

We infer galaxy absolute magnitudes using \textsc{Lephare}\footnote{\url{https://www.cfht.hawaii.edu/~arnouts/LEPHARE/lephare.html}} \citep{Arnouts2011Lephare}, assuming the dust extinction law from \citep{Calzetti1994} and the stellar population synthesis model from \citet{BruzualCharlot2003}. We correct our magnitudes to $z=0$; the K-correction is provided by \textsc{Lephare} and the correction for the evolution of the stellar populations ($e-$correction) is computed with the python package \textsc{EzGal}\footnote{\url{http://www.baryons.org/ezgal}} \citep{Mancone2012EzGal}, assuming Salpeter initial mass function \citep{Chabrier2003} and a single star formation
burst at $z = 3$. These corrections are based on the magnitudes used to define the colours (\texttt{MAG$\_$GAAP}), which are measured using Gaussian apertures \citep{Kuijken2019DR4}. Although ideal for colour estimates, these underestimate the flux and should not be used to compute the luminosity. For that purpose we correct\footnote{The total flux in the $x$ filter can be computed using $m_x=\texttt{MAG\_AUTO}_r+\texttt{(MAG\_GAAP}_x-\texttt{MAG\_GAAP}_r)$, which implicitly assumes that colour gradients are negligible.} them using the Kron-like \texttt{MAG$\_$AUTO} measured from the $r$-band images by \textsc{SExtractor} \citep{Bertin1996SExtractor}. 

The left panel of Fig.~\ref{fig:mag_histo_dense_and_lum} shows the distribution in apparent magnitude \texttt{MAG}$\_$\texttt{AUTO} for galaxies in the \dense\ and \lum\ samples for which shapes were determined by \lensfit\ or \textsc{DEIMOS}. In Sect.~\ref{sec:shape_measurements} we describe the two shape measurement methods and explain the difference in their number counts. We note that the LRGs are much brighter than the limiting magnitude of KiDS in the $r$-band. The corresponding distributions in absolute magnitude in the rest-frame $r$ filter, K+$e$ corrected to $z=0$, are presented in the right panel of Fig.~\ref{fig:mag_histo_dense_and_lum}. This shows that the \dense\ sample overlaps somewhat with the \lum\ sample in terms of luminosity, as a consequence of the photometric redshift uncertainty\footnote{The selection through the redshift-dependent apparent magnitude cut results in an overlap in apparent magnitudes of the \dense\ and \lum\ samples. Because the cut is redshift-dependent, this implies a threshold in luminosity: In the case of perfect redshifts, this would result in a disjoint sample, because we removed the galaxies from the \dense\ sample that overlap with the \lum\ one. The photometric redshift uncertainty, however, assigns to galaxies with the same apparent magnitude different luminosities, and thus a portion of the \dense\ sample extends above the luminosity threshold of the \lum\ sample.}.

\subsection{Satellite galaxy fraction estimation} \label{subsec:satellite_fraction}

Observations suggest that satellite galaxies are only weakly aligned \citep[see e.g.][for recent constraints]{Georgiou2019b} and thus suppress the IA signal at large scales. We do not take this into account in our analysis but provide here an estimate of the fraction of satellites we expect in our samples. Such information will be useful for future modelling studies.

We use the publicly available \texttt{G3GGal} and \texttt{G3GFoFGroup} catalogues \citep[][]{Robotham2011GAMAFoF} from the GAMA survey \citep[][]{Driver2009, Driver2011, Driver2015}. Since KiDS overlaps with GAMA, these catalogues provide group information for a subset of our galaxies, obtained with a Friends-of-Friends algorithm. We cross-match our LRG samples with the \texttt{G3GGal} catalogue and select galaxies with $z<0.21$ ($z<0.32$), which provide a roughly volume-complete match to the \dense\ (\lum) sample. With the information in both group catalogues,  we identify both the brightest group galaxies and ungrouped galaxies as centrals, and the rest as satellites. With this strategy, we obtain $f_\mathrm{sat}=0.34$ for our \texttt{dense}$\times$GAMA sample and $f_\mathrm{sat}=0.23$ for the \texttt{luminous}$\times$GAMA\footnote{These estimates refer to the full samples, but should be representative for the shape samples as well.}. Since our samples are selected to resemble the same galaxy populations at different redshifts, these estimates should be fairly representative beyond the redshift range probed by our direct comparison.

\section{Shape measurements}\label{sec:shape_measurements}

In addition to precise redshifts, a successful IA measurement requires accurate shape measurements. In this work, we compare two different algorithms, \textsc{DEIMOS} and \lensfit\,  both in terms of their ability to recover reliable ellipticity measurements and the resulting IA signal. Exploring the dependence of the IA signal on the shape measurement algorithm is important if one aims to provide informative priors to lensing studies \citep{Singh2016}. Both algorithms have been used to analyse KiDS data: \textsc{DEIMOS} to provide the shape catalogue \citep{Georgiou2019} for a number of IA studies,
while \lensfit\ was used for cosmic shear analyses \citep[see][for the most recent shape measurements]{Giblin2021}.

\subsection{DEIMOS}\label{subsec:DEIMOS}

\textsc{DEIMOS} \citep{Melchior2011} is a moment-based shape measurement algorithm designed to measure the
moments of the surface brightness distribution from an image, which are subsequently used to estimate the ellipticity. The main features of \textsc{DEIMOS} are its rigorous treatment of the PSF moments to arbitrary order, the lack of model assumptions and the flexibility in changing the size of the weight function so that it is possible to assign more importance to different parts of a galaxy while performing the shape measurement (bulge or outskirts). 

The unweighted moments of the surface brightness $G(\vec{x})$ are defined as
\begin{equation}
	\tens{Q}_{ij} \equiv \{ G\}_{ij} = \int G(\vec{x})\, x^{i} y^{j} \, {\rm{d}}x\, {\rm{d}}y \ ,
	\label{eq:unweighted_moments}
\end{equation}
where $(x,y)$ are the Cartesian coordinates with origin at the galaxy's centroid. The complex ellipticity is then defined in terms of the second-order moments as
\begin{equation}
	\label{eq:ellipticity}
	\epsilon \equiv \epsilon_{1} + \rm{i}\epsilon_{2} = \frac{ \tens{Q}_{20} - \tens{Q}_{02} + 2\rm{i}\,\tens{Q}_{11} }{ \tens{Q}_{20} + \tens{Q}_{02} + 2\, \sqrt{\tens{Q}_{20}\, \tens{Q}_{02} - \tens{Q}_{11}^{2}} } \ .
\end{equation}

In practice, unweighted moments cannot be used because of noise in the images, and weighted moments have to be employed instead. We will return to this issue later. Moreover, the galaxy images are smeared and distorted by the atmospheric blurring and the telescope optics, so that the observed image, $G^{\vec{*}}$, is convolved with the PSF kernel $P(\vec{x})$,
\begin{equation}
	G^{\vec{*}}(\vec{x}) = \int G(\vec{x}')\, P(\vec{x} - \vec{x}')\, {\rm{d}}\vec{x}' \ .
\end{equation}

The \textsc{DEIMOS} algorithm estimates the unweighted moments by correcting the 
observed weighted moments of the galaxy surface brightness for the convolution by the PSF. The underlying mathematical framework is a deconvolution in moment space. In order to measure the moments in Eq. (\ref{eq:unweighted_moments}) we then need to deconvolve them. This can easily be achieved in Fourier space, where the convolution becomes a product. Using the Cauchy product, we can write \citep{Melchior2011}: 
\begin{equation} \label{eq:G*_ij}
	\{ G^{*} \}_{ij} = \sum^{i}_{k} \sum^{j}_{l} \, \begin{pmatrix} \, i\, \\ k \end{pmatrix} \begin{pmatrix} j \\ \, l\, \end{pmatrix} \, \{ G \}_{kl} \{P\}_{i-k,j-l} \ ,
\end{equation}
which shows that the ($i+j$)-order convolved moments are determined by the same- or lower-order moments of the galaxy and the PSF kernel. The deconvolution procedure to estimate the galaxy moments is to invert the above hierarchical system of equations, starting from the zeroth order.

As mentioned above, it is necessary to introduce a weight function to avoid noise dominating the second-order moments outside the galaxy light profile. In this work, we adopt an elliptical Gaussian weight function with size $r_\mathrm{wf} = r_\mathrm{iso}$, where $r_\mathrm{iso}$ is the isophotal radius, defined as $r_\mathrm{iso} = \sqrt{A_\mathrm{iso}/\pi}$, following \citet[][]{Georgiou2019}. 
The area $A_{\rm{iso}}$ of the galaxy's isophote is computed using the \texttt{ISOAREA$\_$IMAGE} by \textsc{SExtractor} \citep{Bertin1996SExtractor}.
The shape measurement procedure is the same as described in \citet{Georgiou2019} and we point the interested reader to their Section 2 for a detailed description of the algorithm. In Appendix~\ref{A:mbias_calibration} we report our analysis of the measured shape bias for different setups, which led to our final choice reported above. 

Using \textsc{DEIMOS}, we successfully measure the shapes of 96\,863 galaxies from the \lum\ sample, $\sim 83 \%$ of the corresponding density sample, and 152\,832 shapes from the \dense\ sample, roughly $\sim 88 \%$ of its density sample. The shape measurements mainly fail\footnote{We only considered shapes with \texttt{flag$\_$\textsc{DEIMOS}==0000}, corresponding to measurements that do not raise any flag \citep[see ][]{Georgiou2019}.} for the faintest galaxies in the sample.

\subsection{\lensfit}\label{subsec:lensfit}

The second shape catalogue is obtained using the self-calibrating version of \lensfit\ \citep{Miller2013}, described in more detail in \cite{FenechConti2017}. It is a likelihood-based model-fitting method that fits a PSF-convolved two-component bulge and disk galaxy model. This is applied simultaneously to the multiple exposures in the KiDS-1000 $r$-band imaging, to get an ellipticity estimate for each galaxy. 

\lensfit\ provides shapes for 84\,785 galaxies from the \lum\ sample ($72\%$ of the density sample), and for 121\,500 galaxies from the \dense\ sample ($70\%$ of the density sample). The lower completeness with respect to \textsc{DEIMOS} is largely explained by the fact that \lensfit\ has been optimised for cosmic shear studies, where the signal is maximised for high-redshift galaxies, which are typically small and faint.  Whilst \lensfit\ could determine ellipticity measurements for the large bright galaxies with $\texttt{MAG\_AUTO}<20$, this model-fitting algorithm becomes prohibitively slow given the large number of pixels that these bright galaxies span. Therefore, the \lensfit\ catalogue only contains galaxies fainter than $\texttt{MAG\_AUTO}>20$  (hence the sharp cut-off in apparent magnitude in Fig.~\ref{fig:mag_histo_dense_and_lum}). It performs better than \textsc{DEIMOS} for relatively faint and low S/N galaxies. As these are preferentially found at higher redshifts, this also explains the different redshift distributions, as illustrated in Fig.~\ref{fig:z_histo_shapes}.

\subsection{Image simulations}\label{subsec:image_simulations}

\begin{figure*}
\centering
\subfloat[]{{
\includegraphics[width=0.9\columnwidth]{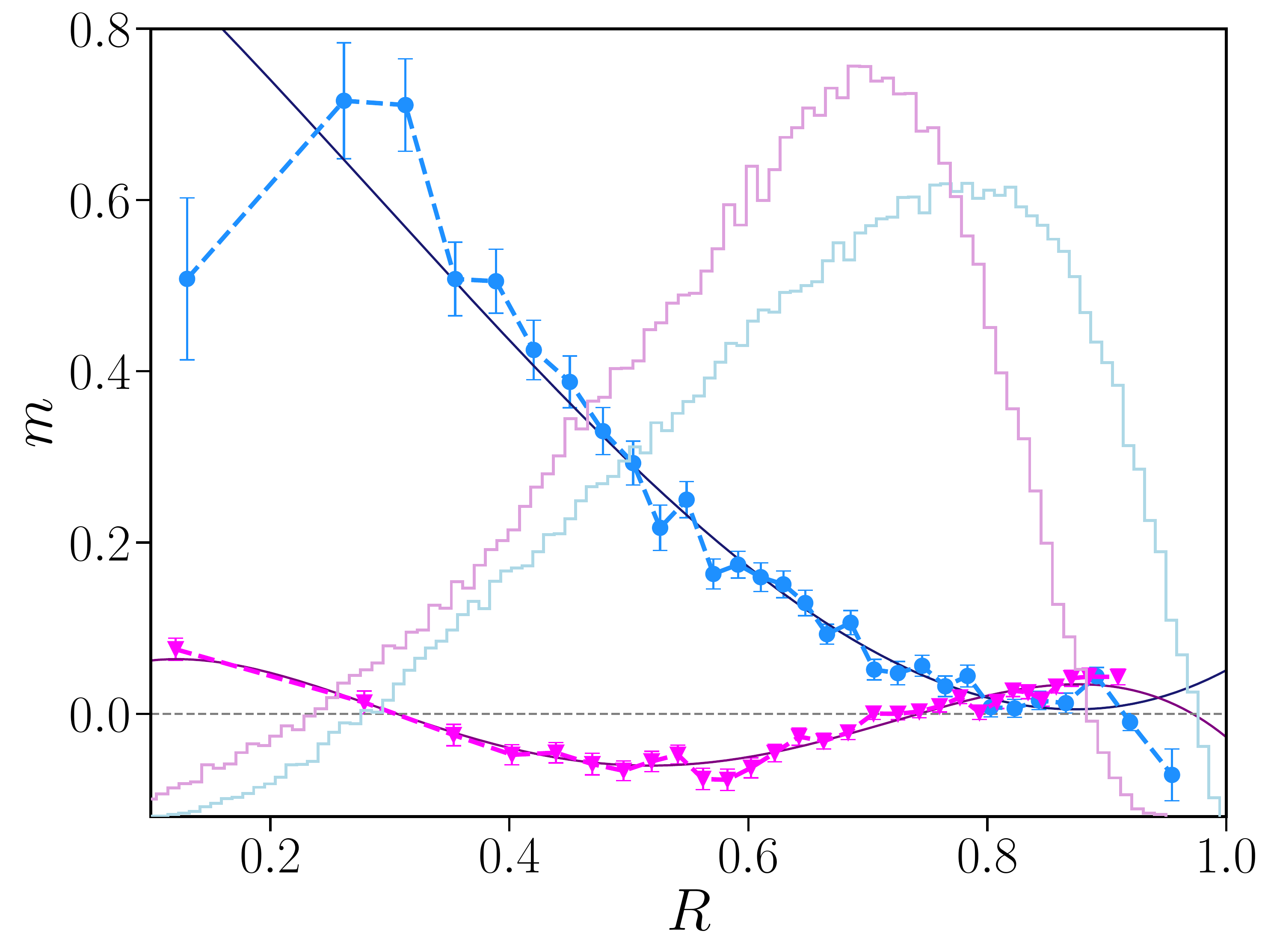}
}}
\qquad
\subfloat[]{{
\includegraphics[width=0.9\columnwidth]{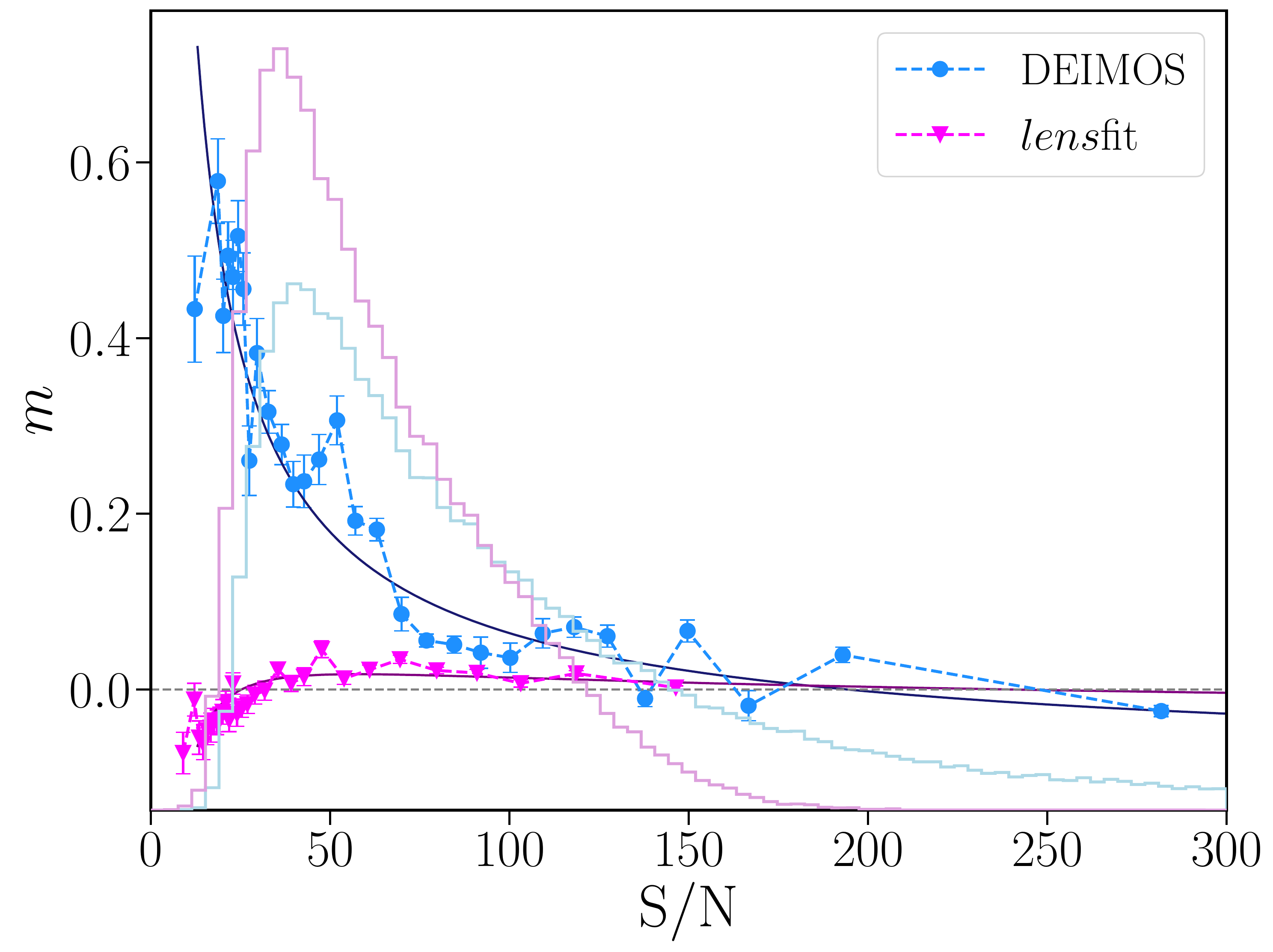}}}
\caption{Average multiplicative bias, $m = (m_{\epsilon_1}+m_{\epsilon_2})/2$, as a function of (a) the galaxy resolution, $R$, and (b) the signal-to-noise ratio, S/N. Each point is measured on the same number of simulated galaxies and the error bars are estimated using bootstraps. For a comparison we also display in the background the weighted distribution of the two definitions of $R$ and the S/N in the real data for the \dense\ shape samples (pink: $lens$fit; blue: \textsc{DEIMOS}). The solid lines show the polynomial fit to $m(R)$ and $m(\mathrm{S/N})$, which guided the construction of the two-dimensional bias surface.
}
\label{fig:mbias_gals_par}
\end{figure*}

We want to measure the shapes of galaxies from images that are corrupted by noise and blurred by the atmosphere and telescope optics. These bias the inferred shapes and thus need to be carefully corrected for. Although both
\textsc{DEIMOS} and \lensfit\ are designed to do so, residual biases remain. These can be expressed as \citep{Heymans2006}
\begin{equation}
    \epsilon_i^\mathrm{obs} = (1+m_i) \epsilon_i^\mathrm{true} + c_i \ ,
\end{equation}
with $i\in\{1,2\}$ the ellipticity components introduced in \ref{eq:ellipticity}. Here $\epsilon_i^\mathrm{true}$ is the true ellipticity, while $\epsilon_i^\mathrm{obs}$ is the output of the shape measurement algorithm; $m_i$ is the multiplicative bias and $c_i$ is the additive bias. Differently from what is done in lensing studies \citep[e.g.][]{Kannawadi2019}, here we calibrate the ellipticity rather than the shear.
Our aim is to determine the biases in our shape measurements using realistic image simulations, with a precision that is better than the statistical error on our IA signal. 

We stress that although it is important to start with an algorithm that does not lead to a large bias in the first place, what matters the most is to calibrate the residual bias on realistic image simulations in order to properly account for galaxy blending and the different observing conditions \citep{Hoekstra2017, Kannawadi2019, Samuroff2018IM3SHAPE, MacCrann2020}. We use dedicated image simulations generated with the COllege pipeline \citep[COSMOS-like lensing emulation of ground experiments; ][]{Kannawadi2019}. These simulations reproduce the observations from the Cosmic Evolution Survey \citep[COSMOS,][]{Scoville2007}, for which we have both KiDS imaging (KiDS-COSMOS) and deeper images from the \textit{Hubble} Space Telescope (HST). We use the HST observations to generate our input catalogue and simulate the KiDS observations by varying the observation conditions. Under the assumption that COSMOS is representative of our galaxy sample (in practice we only require that it covers the signal-to-noise (S/N) and size parameter space, while we do not need the galaxy distributions to match) we study the $m-$bias properties of the LRGs in our KiDS-COSMOS field and use the bias model obtained from this set of galaxies to calibrate our full sample.

The image simulations used in this work differ slightly from those presented in \citet[]{Kannawadi2019} because we require a larger number of simulated LRGs for our calibration. To achieve this, we adopt the ZEST catalogue \citep[Zurich Estimator of Structural Type;][]{Scarlata2007, Sargent2007} for the input galaxy parameters. We generated 52 KiDS-like images by varying the observing conditions and rotating the galaxies. We used 13 different PSF sets and four rotations per each image. Since our underlying galaxy selection is identical for both the \lensfit\ and \textsc{DEIMOS} shape catalogues, we employed the same suite of simulations for both calibrations.

The shape measurement bias depends on the size, S/N, radial surface brightness profile and ellipticity of the galaxy, as well as the observing conditions. Of these, the size and S/N are the most relevant, and we use these to capture the dependence of the bias for our set of simulated galaxies. Rather than the intrinsic size of the galaxy, we use a proxy for how well it is resolved: $R$ quantifies the relative size of the PSF compared to the size of the galaxy. Here, we adopt two slightly different definitions, depending on the shape algorithm employed. For \textsc{DEIMOS} we use
\begin{equation}
    R^\mathrm{DEIMOS} = 1 - \frac{T^\mathrm{PSF}}{T^\mathrm{gal}} \ ,
\end{equation}
where $T^\mathrm{PSF} = \tens{Q}^\mathrm{PSF}_{20} + \tens{Q}^\mathrm{PSF}_{02}$ and $T^\mathrm{gal} = \tens{Q}^{\vec{*} 
\mathrm{gal}}_{20} + \tens{Q}^{\vec{*} \mathrm{gal}}_{02}$, where $\tens{Q}_{ij}^{\vec{*}\mathrm{gal}}$ are the unweighted moments of the PSF-convolved surface brightness profile (see Eqs. \ref{eq:G*_ij} and \ref{eq:unweighted_moments}). 
In the case of $lens$fit we use
\begin{equation}
    R^{lens\mathrm{fit}} = 1 -  \frac{r^2_\mathrm{PSF}}{\left( r^2_\mathrm{ab} + r^2_\mathrm{PSF} \right)} \ ,
\end{equation}
where $r^2_\mathrm{PSF} = \sqrt{P_{11} P_{22} - P_{12}^2}$ and $r_\mathrm{ab} = r_{\rm e} \sqrt{q}$. Here, $P_{ij}$ are the $lens$fit PSF weighted quadrupole moments \citep[see Eq. (2) in ][]{Giblin2021}, measured with a circular Gaussian function of size $2.5$ pixels; $r_{\rm e}$ is the half-light radius measured along the major axis of the best-fit elliptical profile by $lens$fit, which is an estimate of the true galaxy size before PSF-convolution, while $q$ is the axis ratio, such that $r_\mathrm{ab}$ is the azimuthally averaged size of the galaxy. As we can see, $R$ can in practice only assume values between 0 and 1, where 1 corresponds to galaxies with sizes that are much larger than the PSF.

We evaluate the multiplicative bias $m$ in bins of S/N and $R$ that contain an equal number of galaxies and the error bars are computed using 500 bootstrap realisations. The resulting biases are presented in Fig. \ref{fig:mbias_gals_par} for both \lensfit\ and \textsc{DEIMOS}. 
We find that the two components $\epsilon_{1,2}$ show similar dependencies, and we, therefore, calibrate the bias for the two components jointly. The additive bias for both components is consistent with zero, and thus we do not consider it further in our calibration. 

For both $m(\mathrm{S/N})$ and $m(R)$, we find that \lensfit\ has a small bias and thus also our correction is small; in general, it performs better than \textsc{DEIMOS} for poorly resolved galaxies and low S/N. It is, however, prohibitively slow when measuring shapes for large galaxies, limiting the lensfit sample to galaxies with $m_r > 20$. In contrast, \textsc{DEIMOS} shows a large bias for low values of $R$: the galaxy size correlates with its ellipticity, and we find that removing the highly elliptical galaxies significantly reduces the bias. However, once we calibrate the shapes of those galaxies, we recover a very similar signal for the full shape sample and the one cut in ellipticity. Similarly, we have also tested that adding inverse-variance weights to account for these noisy galaxies does not significantly improve our signal. This motivates our choice to keep all galaxies in our sample and not to introduce additional weighting; we assume that the measurements are dominated by shape noise only.

We can see that $m(R)$ for both \textsc{DEIMOS} and \lensfit\ is well described by a polynomial curve, which we truncate at degree 3 and 4, respectively, while $m(\mathrm{S/N})$ is well described by the expansion: $d(\mathrm{S/N}) = d_1/\sqrt{\mathrm{S/N}} + d_2/(\mathrm{S/N})$. We combine the two individual bias dependencies into a single bias surface as detailed in Appendix~\ref{A:mbias_calibration}. The specific functional forms for the two shape methods differ to better adapt the surface to our observed bias. We use these empirical relations to infer the $m$-bias associated with each galaxy, given its S/N and $R$. 

To ensure that our empirical correction performs well on our sample, we select sets of galaxies from the image simulations that resemble our LRG samples by reproducing the observed distributions in S/N and $R$. We measure the residual biases for these samples, defined as the difference in the estimated $m$-bias (inferred using our model for the bias) and the bias
measured directly from the simulations for the given set of galaxies. For the \textsc{DEIMOS} shape method, we find an average residual of $-0.002 \pm 0.007$
for the \dense-like sample, while this is $-0.002 \pm 0.008$ for the \lum-like sample. Similarly, in the case of \lensfit\, the residuals for the \lum-like and \dense-like galaxies are, respectively, $-0.0014 \pm 0.0013$ and $-0.0019 \pm 0.0020$. As we will see later, this is much smaller than the uncertainty in the IA measurements: the average bias introduced by the shape measurement process is subdominant and does not affect our best estimate of the IA amplitude.

\begin{figure}
\centering
\includegraphics[width=\columnwidth]{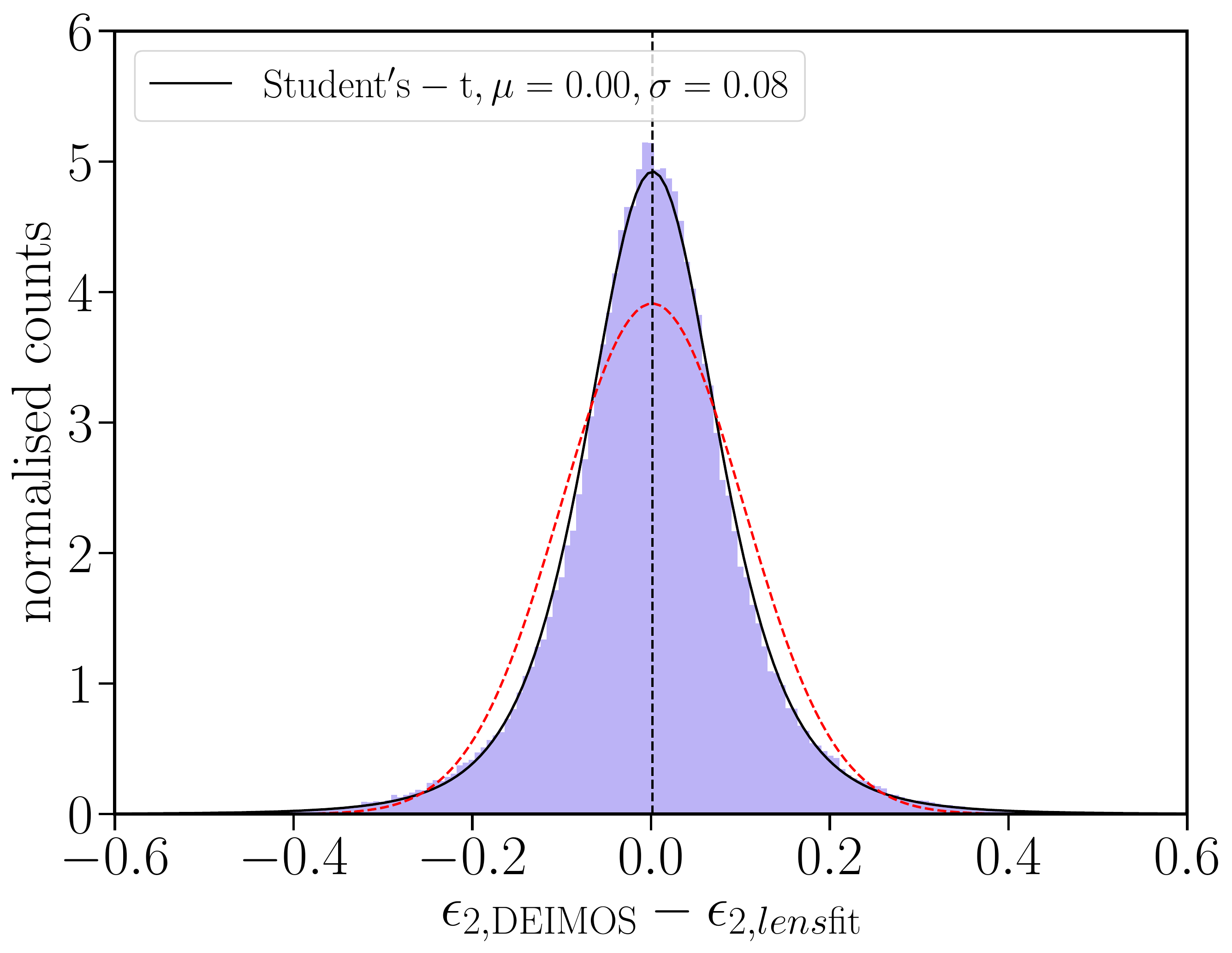}
\caption{Histogram of the difference of the $\epsilon_1$ component of the ellipticity measured by the two shape measurement algorithms, \lensfit\ and \textsc{DEIMOS}, on a common sub-sample of galaxies, after applying the $m-$bias correction as described in the text. The $\epsilon_2$ component shows the same behaviour. The distribution is more peaked than a Gaussian (red dashed line) and is best described by a Student's $t-$distribution with $\nu=4.3$, and a width $\sigma=0.08$ with zero mean (black solid line).}
\label{fig:e1_lensfit_vs_DEIMOS}
\end{figure}

The LRGs are relatively bright and we thus expect the shape measurements to be shape noise-dominated. This also implies that the \textsc{DEIMOS} and \lensfit\ measurements are correlated. To quantify this, we show the distribution of the difference between the $m-$corrected ellipticities measured by the two algorithms in Fig. \ref{fig:e1_lensfit_vs_DEIMOS}. The distribution is more peaked than a Gaussian, and well described by a Student's $t-$distribution centred on zero, with $\nu=4.30$ (degrees of freedom) and with scale parameter $\sigma=0.08$. This is to be compared to the intrinsic ellipticity of galaxies, which is about $\epsilon_\mathrm{rms} = 0.12$ based on \textsc{DEIMOS} measurements for galaxies with apparent magnitude $m_r<20$. It is interesting to note that our sample is considerably rounder than a typical cosmic shear sample, as expected for an LRG sample \citep[see for example][]{vanUitert2012}; this implies that it might be affected differently by a weighting scheme in a lensing analysis. The differences between the \textsc{DEIMOS} and \lensfit\ measurements are caused by differences in how each method deals with noise in the images.

\section{Correlation function measurements} \label{sec:correlation_function_measurements}

We measure the IA signal using the two-points statistic $\wgp$, defined as the projection along the line-of-sight of the cross-correlation between galaxy positions and galaxy shapes. It measures the tendency of galaxies to point in the direction of another galaxy as a function of their comoving transverse separation, $r_{\rm p}$, and comoving line-of-sight separation, $\Pi$. To quantify the alignment signal in our data, we employ the estimator presented in \citet{Mandelbaum2006a}\footnote{Instead of normalising by $R_{\rm S} R_{\rm D}$, we actually normalise by the density - randoms vs. shapes pair count, $R_{\rm D} D_{\rm S}$. This significantly speeds up the computation and has been tested to have negligible impact \citep{Johnston2019}.},
\begin{equation}
    \hat{\xi}_{g+} (r_{\rm p},\Pi) = \frac{S_+D-S_+R_{\rm D}}{R_{\rm S} R_{\rm D}},
\end{equation}
where $R_{\rm D}$ and $R_{\rm S}$ are catalogues of random points designed to reproduce the galaxy distribution of the density and shape samples, respectively. We indicate with $D$ the density sample that provides the galaxy positions, while $S_+$ is the shape sample, such that the quantity
\begin{equation}
    S_+D = \sum_{i \neq j} \gamma_+(i|j),
\end{equation}
gives us the tangential shear component of the galaxy pair $(i,j)$, $\gamma_+(i|j)$, where $i$ is extracted from the shape sample and $j$ from the density sample. $\gamma_+$, in turn, is defined as
\begin{equation}
    \gamma_{+} (i|j) = \frac{1}{\mathcal{R}} \Re \left[\epsilon_i \exp(-2i\phi_{ij} )\right],
\end{equation}
where $\Re$ denotes the real part; $\epsilon_i$ is the complex ellipticity associated with the galaxy $i$, $\epsilon_i = \epsilon_{1,i} + i \epsilon_{2,i}$, whose components 1,2 are measured by the shape measurement algorithms presented in Sect. \ref{sec:shape_measurements}; $\phi_{ij}$ is the polar angle of the vector that connects the galaxy pair; $\mathcal{R}=\partial\epsilon/\partial\gamma$ is the shear responsivity and it quantifies by how much the ellipticity changes when a shear is applied: for an ensemble of sources,  $\mathcal{R} = 1 - \epsilon^2_\mathrm{rms}$.

The galaxy clustering signal is computed with the standard estimator \citep{Landy&Szalay1993},
\begin{equation}
    \hat{\xi}_\mathrm{gg} (r_{\rm p}, \Pi) = \frac{DD - 2 DR_D - R_D R_D}{R_D R_D} \ .
\end{equation}{}

To measure our clustering and IA signals, we use uniform random samples that reproduce the KiDS footprint, accounting for the masked regions; to these we assign redshifts randomly extracted from the galaxy unconditional photometric redshift distributions. For each sample, we construct the random sample to match their redshift distribution.

To account for the spatial variation in the survey systematics, we apply weights to the galaxies when computing the signal, as discussed in \citet[][]{Vakili2020}. These weights are designed to remove the systematic-induced variation in the galaxy number density across the survey footprint. For a detailed discussion of how the weights are generated and tested, we refer to Sect.~4 in \citet[][]{Vakili2020}. To capture the variation in the survey systematics along the line-of-sight, we split each sample into three redshift bins and assign the weights to those sub-samples. We tested that this procedure does not induce a correlation between the galaxy weights and the redshifts themselves. We have also verified that the impact of the weights is very small and can be neglected when considering the split in luminosity of the samples (see Sect.~\ref{subsec:luminosity_dependence}). We apply such weights to both the density and shape samples.

In this work, we measure the clustering and IA signals using an updated version of the pipeline presented in \citet[]{Johnston2019}, which makes use of the publicly available software \textsc{Treecorr} \citep[][]{Jarvis2004treecorr}\footnote{\url{https://github.com/rmjarvis/TreeCorr}} for clustering correlations. \xigp and \xigg are then projected by integrating over the line-of-sight component of the comoving separation, $\Pi$,
\begin{equation} \label{eq:w_estimator}
    \hat{w}_{\mathrm{g} i} (r_{\rm p}) = \int_{- \Pi_\mathrm{max}}^{\Pi_\mathrm{max}} \dd \Pi \ \hat{\xi}_{\mathrm{g} i}(r_{\rm p}, \Pi) \quad i=\{+, g \} \ .
\end{equation}

The largest scales probed in this analysis are limited by the effective survey area ($\sim777$ deg$^2$). We set a maximum transverse separation of $60 \mpch$ and measure the signal in 10 logarithmically spaced bins, from $r_{\rm p, min}=0.2 \mpch$. 

We perform the measurements for three different setups: we adopt $\Pi_\mathrm{max}=120 \mpch$ as the fiducial case, but repeat the analysis for $\Pi_\mathrm{max}= 90 \mpch$ and $\Pi_\mathrm{max}= 180 \mpch$ (see Appendix~\ref{A:systematic_tests}). We always bin our galaxies in equally spaced bins with $\Delta \Pi = 10 \mpch$. We observe an extended signal to $\Pi>180 \mpch$, but the signal is comparable to the noise at those distances.

Our choice of $\Pi_\mathrm{max}$ is conservative since the uncertainties in the photometric redshifts are $\sigma_z<0.02(1+z)$ for both the \dense and \lum\ samples \citep{Vakili2020}, and if we choose $\Pi_\mathrm{max}$ based on the $1\sigma$ uncertainty in the photometric redshifts \citep{Joachimi2011b}, we could potentially reduce $\Pi_\mathrm{max}$  to $70 \mpch$. However, this might be too optimistic given that the error on $\sigma_z$ increases with redshift.
The choice of $\Pi_\mathrm{max}$ is motivated by two opposite necessities: to maximise the S/N, we want to minimise the amount of signal that we discard, whilst we also want to avoid adding uncorrelated pairs that would increase the noise. To find the best balance, we calculate the S/N of our signal as a function of $(r_{\rm p}, \Pi)$ by dividing the measured $w_\mathrm{gg}(r_{\rm p},\Pi)$  by the root-diagonal of the jackknife covariance. 
We truncate at $\Pi_\mathrm{max}$ based on the 10 $\sigma$ detection, which roughly corresponds to $\Pi_\mathrm{max} = 120 \mpch$. In addition to these considerations, there is a further motivation to limit the integral to modest line-of-sight separations: as discussed in Appendix~\ref{A:lensing_contamination}, the contamination from galaxy-galaxy lensing has a shallower dependence on the line-of-sight separation; as we move along the $\Pi$ direction, we see an increase in the contamination with a mild increase in the IA signal, until lensing dominates.

The error bars are computed via a delete-one jackknife re-sampling of the observed volume. The covariance matrix is constructed as 
\begin{equation}
    \mathrm{Cov}_\mathrm{jack.} = \frac{N-1}{N} \sum_{\alpha = 1}^{N} (w^{\alpha} - \bar{w}) (w^{\alpha} - \bar{w})^\top ,  
\end{equation}
where $w^{\alpha}$ is the signal measured from jackknife sample $\alpha$, while $\bar{w}$ is the average over $N$ samples; $\top$ denotes the transpose of the vector.

The number of regions $N$ is ultimately set by the size of the survey and the scales we aim to probe. A maximum value of $r_{\rm p}=60 \mpch$ corresponds to an angular separation of $\sim 8$ degrees (\dense\ sample) and $\sim 6 $ degrees (\lum\ sample) at the lowest redshifts probed in the analysis. However, to increase the number of jackknife regions, we decide to set the minimum angular scale to 5 degrees, which strictly satisfies our requirement only for $z\gtrsim 0.2$. This is motivated by the fact that the majority of our galaxies are at high redshift and hence only $\lesssim 5\%$ of our galaxies have unreliable error estimates in the last $r_{\rm p}-$bin. The total number of jackknife regions that we are able to obtain for our samples is $N = 37$. 
We correct our inverse covariance matrices, which enter into our likelihood estimations,  as recommended in \citet[][]{Hartlap2007}: because of the presence of noise, the inverse of a covariance matrix obtained from a finite number of jackknife (or bootstrap) realisations is a biased estimator of the true inverse covariance matrix.

\section{Modelling}
\label{sec:theoretical_framework}

The linear alignment model \citep{Catelan2001, Hirata2004} predicts a linear relation between the contribution to the shear induced by IA and the quadrupole of the gravitational field responsible of the tidal effect. This can be expressed as
\begin{equation}\label{eq:intrinsic_shear_equation}
    \gamma^{\rm I} = (\gamma^{\rm I}_+, \gamma^{\rm I}_\times) = -\frac{C_1}{4 \pi G} (\partial_x^2 + \partial_y^2, \partial_x \partial_y) \Phi_{\rm p} \ , 
\end{equation}
where the partial derivatives are with respect to comoving coordinates and provide the tangential and cross components of the shear with respect to the $x$-axis; $\Phi_{\rm p}$ is the gravitational potential at the moment of galaxy formation, assumed to take place during the matter-dominated era \citep{Catelan2001}; $C_1$ is a normalisation constant and $G$ is the gravitational constant. 

Using Eq. (\ref{eq:intrinsic_shear_equation}), by correlating the intrinsic shear with itself or with the matter density field $\delta$, we can construct the relevant equations for the IA correlation functions \citep{Hirata2004}. In Fourier space, the matter density-shear power spectrum $(\delta {\rm I})$ becomes
\begin{equation}\label{eq:LA_deltaI}
    P^{\mathrm{LA}}_{\delta \mathrm{I}}(k,z) = A_\mathrm{IA} C_1 \rho_{\rm c} \frac{\Omega_m}{D(z)} P_{\delta \delta}^{\mathrm{lin}}(k,z) \ .
\end{equation}
Here, $D(z)$ is the linear growth factor, normalised to unity at $z=0$, $\rho_{\rm c}$ is the critical density of the Universe today, and $P_{\delta \delta}^{\mathrm{lin}}$ is the linear matter power spectrum. We set $C_1=5\times10^{-14} h^{-2} M_\odot^{-1} \mathrm{Mpc}^3$  based on the IA amplitude measured at low redshifts using SuperCOSMOS \citep{Brown2002}, which is the standard normalisation for IA power spectra. 

Galaxies are biased tracers of the matter density field, and at large scales this relation is linear, $\delta_{\rm g} \sim b_{\rm g} \delta$. We can thus relate the galaxy position--intrinsic shear power spectrum to the matter density--intrinsic shear power spectrum via the galaxy bias $b_{\rm g}$:
\begin{equation}\label{eq:LA_g+}
    P^{\mathrm{LA}}_{\mathrm{g I}}(k,z) = b_{\rm g}  P^{\mathrm{LA}}_{\delta \mathrm{I}}(k,z) \ ,
\end{equation}
which is the power spectrum of interest for our analysis.

A successful modification of the LA model replaces the linear matter power spectrum in Eq.~\ref{eq:LA_deltaI} with the non-linear one, to account for the non-linearities arising at intermediate scales \citep{BridleKing2007}. This so-called NLA model was succesfully employed in a number of studies \citep[e.g.][]{Blazek2011, Joachimi2011b} and here we follow the same approach to model our signal. More sophisticated treatments of the IA signal, which include the modelling of the mildly or fully non-linear scales, have been developed in the last decade \citep[][]{SchneiderBridle2010, Blazek2019TATT, fortuna2020halo}, but given the scales probed in our analysis (see Sect.~\ref{subsec:likelihoods}) and the homogeneous characteristics of the galaxy population studied, the NLA model provides a sufficient description for this work. Unless stated otherwise, in the following we always assume the NLA model as our reference choice. To generate the linear matter power spectrum we use \textsc{CAMB}\footnote{https://camb.info} \citep{Lewis2000, Lewis:2002ah}, while the non-linear modifications are computed using \textsc{Halofit} \citep{Smith2002} with the implementation presented in \citet{ Takahashi2012halofit}. In the rest of the paper, we simply refer to the non-linear matter power spectrum as $P_{\delta \delta}(k,z)$.

\subsection{Incorporating the photometric redshift uncertainty into the model}

The use of photometric redshifts results in an uncertainty in the estimated distance of the galaxies, which has to be included in the model. In particular, if we express the correlation function $\xi_\mathrm{gI}$ in terms of the two components of the galaxy separation vector $\mathbf{r}$, $(r_{\rm p}, \Pi)$, we can map the redshift probability distribution into the probability that the true values of $r_{\rm p}$ and $\Pi$ correspond to their photometric estimates. Here, we follow the approach derived in \citet[]{Joachimi2011b} and use their approximated expression,
\begin{equation} \label{eq:xi_gI}
    \xi^\mathrm{ph}_\mathrm{gI}(\bar{r}_{\rm p}, \bar{\Pi}, \bar{z}_{\rm m}) = \int \frac{\dd \ell \ell }{2 \pi} J_2 \left(\ell \theta (\bar{r}_{\rm p}, \bar{z}_{\rm m}) \right)C_\mathrm{gI} \left( \ell; \bar{z}_1 (\bar{z}_{\rm m}, \bar{\Pi}), \bar{z}_2(\bar{z}_{\rm m}, \bar{\Pi}) \right) \ .
\end{equation}{}

The observables are: $\bar{z}_1$ and $\bar{z}_2$, the photometric redshift estimates of the pair of galaxies for which we are measuring the correlation, and their angular separation $\theta$. These can be related to $(\bar{r}_{\rm p}, \bar{\Pi}, \bar{z}_{\rm m})$, through the approximate relations
\begin{align} \label{eq:coordinate_transformation}
    z_{\rm m} & = \frac{1}{2} (z_1 + z_2) \ , \\
    r_{\rm p} & \approx \theta \chi (z_{\rm m}) \ , \\
    \Pi & \approx \frac{c}{H(z_{\rm m})} (z_2 - z_1) \ , 
\end{align}{}
where $\chi(z_{\rm m})$ and $H(z_{\rm m})$ are, respectively, the comoving distance and the Hubble parameter at redshift $z_m$, and $c$ is the speed of light.

The conditional redshift probability distributions are incorporated into the angular power spectrum $C_\mathrm{gI}$, which can be expressed in terms of the three-dimensional power spectrum $P_\mathrm{gI}(k,z)$,
\begin{multline}\label{eq:C_gI}
    C_\mathrm{gI}(\ell, \bar{z}_1, \bar{z}_2) = \int_0^{\chi_\mathrm{hor}} \dd \chi' \frac{p_n(\chi'|\chi(\bar{z}_1)) p_{\epsilon}(\chi'|\chi(\bar{z}_2))}{\chi'^2} \\  \times P_\mathrm{gI} \left( \frac{\ell + 1/2}{\chi'}, z (\chi') \right)
\end{multline}
where we have implicitly assumed the flat-sky and Limber approximations, and $n$ and  $\epsilon$ indicate the density and shape sample respectively. $p(\chi'|\chi)$ are the conditional comoving distance probability distributions, which are related to the redshift distributions via $p(\chi'|\chi) \dd \chi = p(z|\bar{z}) \dd z$. When computing our predictions, we bin our photometric data and compute the corresponding $p(z|\bar{z}) \equiv p(z_\mathrm{spec}|z_\mathrm{phot})$ per each bin; $z_1$ and $z_2$ in Eq. (\ref{eq:coordinate_transformation}) corresponds to the mean values of the  probability distribution with $z_1$ being the mean of the i-th bin and $z_2$ of the j-th bin. In Appendix \ref{A:redshift_distributions} we show the redshift distributions entering our analysis. We refer the interested reader to appendices A.2 and A.3 in \citet[]{Joachimi2011b} for the full derivation of equation \ref{eq:C_gI}. The exact same formalism can then be applied to the clustering signal, where $C_\mathrm{gI} \to C_\mathrm{gg}$, $J_2 \to J_0$ and the redshift distributions are those corresponding to the density sample.

The projected correlation functions $\wgp$ and $w_{gg}$ can then be obtained as:
\begin{equation}
    w_{\mathrm{g+}}(r_{\rm p}) = \int \dd \bar{\Pi} \ \int \dd z_\mathrm{m} \mathcal{W}(\bar{z}_\mathrm{m})  \xi^\mathrm{ph}_\mathrm{gI} (\bar{r}_{\rm p}, \bar{\Pi}, \bar{z}_\mathrm{m}) 
\end{equation}
and
\begin{equation}
    w_\mathrm{gg}(r_{\rm p}) = \int \dd \bar{\Pi} \ \int \dd z_\mathrm{m} \mathcal{W}(\bar{z}_\mathrm{m})  \xi^\mathrm{ph}_\mathrm{gg} (\bar{r}_{\rm p}, \bar{\Pi}, \bar{z}_\mathrm{m}) \ ,
\end{equation}
where the redshift window function $\mathcal{W}(z)$ is defined as \citep{Mandelbaum2011WiggleZ}:
\begin{equation}
    \mathcal{W}(z) = \frac{p_i(z) p_j(z)}{\chi^2(z) \dd \chi / \dd z} \left[ \int \dd z \frac{p_i(z) p_j(z)}{\chi^2(z) \dd \chi / \dd z} \right]^{-1} \ ,
\end{equation}
where $p_{i,j}(z)$ with ${i,j} \in {S,D}$ are now the unconditional redshift distributions for the shape and density samples, and $\chi(z)$ is the comoving distance to redshift $z$. 


\subsection{Contamination to the signal} \label{sec:contaminations}

All possible two-point correlations between galaxy shapes and positions contribute to the estimator in Eq. (\ref{eq:w_estimator}). Following the notation in \cite{JoachimiBridle2010J}, here we consider: the correlation between the intrinsic shear and the galaxy position (g+), which is the quantity we aim to constrain; but also the correlation between gravitational shear and galaxy position, sourced by the galaxy lensing of a background galaxy by a foreground galaxy (gG); and the apparent modification of the galaxy number counts due to the effect of lensing magnification, which affects both the correlations with the intrinsic shear and the gravitational shear (mI and mG).

Among these effects, galaxy-galaxy lensing is the main  contaminant to our signal. While IA requires physically close galaxies, galaxy-galaxy lensing occurs between galaxies at different redshifts. This implies that the level of contamination depends on our ability to select close pairs of galaxies, which ultimately depends on the photometric redshift precision. 
For this reason, the width and the tails of the redshift distributions play an important role in the amount of contamination. Since our $p(z^\mathrm{spec}|z^\mathrm{phot})$ are quite narrow (see Appendix \ref{A:redshift_distributions}) we do not expect this to be a major effect in our data. Nevertheless, we fully model both lensing and magnification effects, and account for them when interpreting the signal. We note that the sign of the gI and gG terms are opposite, such that adding the lensing to the model allows us to remove its suppressing contribution and capture the true IA signal.

It is convenient to write the various correlations in terms of the projected angular power spectra: indicating with $n$ the density sample (that provides the galaxy positions) and with $\epsilon$ the shape sample, we have
\begin{equation} \label{eq:C_n_eps}
    C^{(ij)}_{n \epsilon}(\ell) =  C^{(ij)}_\mathrm{gI}(\ell) +  C^{(ij)}_\mathrm{gG}(\ell) +  C^{(ij)}_\mathrm{mI}(\ell) +  C^{(ij)}_\mathrm{mG}(\ell) \ ,
\end{equation}
where, in a flat cosmology, these read
\begin{equation}
    C_\mathrm{gG}^{(ij)}(\ell) = b_g \int_0^{\chi_\mathrm{hor}} \dd \chi \frac{p_n^{(i)}(\chi) q_\epsilon^{(j)}(\chi)}{\chi^2} P_{\delta \delta} \left( \frac{\ell + 1/2}{\chi}, \chi \right)\ ,
\end{equation}
\begin{equation} \label{eq:C_mI}
    C^{(ij)}_\mathrm{mI}(\ell) = 2 (\alpha^{(i)} - 1) C^{(ij)}_\mathrm{IG}(\ell),
\end{equation}
and
\begin{equation} \label{eq:C_mG}
    C^{(ij)}_\mathrm{mG}(\ell) = 2 (\alpha^{(i)} - 1) C^{(ij)}_\mathrm{GG}(\ell) \ .
\end{equation}
Here $\alpha^{(i)}$ is the slope of the faint-end logarithmic luminosity function\footnote{Formally, the magnification of the \lensfit\ sample is also affected by the slope of the luminosity function at the bright end of $m_r=20$. We ignore such complexity: we find magnification to be a subdominant effect for the faint distant galaxies, thus the contribution of low-redshift galaxies is expected to be negligible for our analysis.}.
The lensing weight function, $q_X$, $X \in \{n, \epsilon\}$ is defined as
\begin{equation}
    q_X(\chi) = \frac{3 H_0^2 \Omega_m}{2 c^2} \frac{\chi}{a(\chi)} \int_0^{\chi_\mathrm{hor}} \dd \chi' p_X(\chi') \frac{\chi' - \chi}{\chi'} \ .
\end{equation}
$C^{(ij)}_\mathrm{IG}$ is the intrinsic-shear power spectrum. It models the correlation between the shearing of source galaxies by a foreground matter overdensity and the simultaneous IA of galaxies located near that overdensity:
\begin{equation} \label{eq:C_IG}
    C_\mathrm{IG}^{(ij)}(\ell) = \int_0^{\chi_\mathrm{hor}} \dd \chi \frac{p_n^{(i)}(\chi) q_\epsilon^{(j)}(\chi)}{\chi^2} P_{\delta {\rm I}} \left( \frac{\ell + 1/2}{\chi}, \chi \right)\ ;
\end{equation}
$C^{(ij)}_\mathrm{GG}$ is instead defined as:
\begin{equation} \label{eq:C_GG}
    C_\mathrm{GG}^{(ij)}(\ell) = \int_0^{\chi_\mathrm{hor}} \dd \chi \frac{q_n^{(i)}(\chi) q_\epsilon^{(j)}(\chi)}{\chi^2} P_{\delta \delta} \left( \frac{\ell + 1/2}{\chi}, \chi \right)\ .
\end{equation}
We note that with respect to the usual shear power spectrum, we require here that one of the samples refers to the density sample, $n$.

To account for these sources of contamination in the fit, we replace $\xi_\mathrm{gI}$ with $\xi_\mathrm{n \epsilon}$, which can be obtained from Eq. (\ref{eq:C_n_eps}). The prediction for $\xi^\mathrm{obs}$ is then used to constrain the measured signal $\hat{w}_\mathrm{g+}$. In Appendix~\ref{A:lensing_contamination} we expand further on the impact of lensing on our measurements, while in Appendix~\ref{A:magnification_contamination} we describe our strategy to measure the values of $\alpha^{(i)}$ in our data.

\subsection{Likelihoods} 
\label{subsec:likelihoods}

We perform the fits to the data using a Markov Chain Monte Carlo (MCMC)
that samples the multi-dimensional parameter posterior distributions and finds the set of parameters that maximise the likelihood. We assume a Gaussian likelihood of the form $\mathcal{L} \propto \exp(-\chi^2/2)$, where
\begin{equation}
    \chi^2 = \chi^2_{w_{\rm gg}} + \chi^2_{\wgp}
\end{equation}
and we simultaneously fit for the galaxy bias, $b_{\rm g}$ and the IA amplitude, $A_{\rm IA}$.

To correct for the effects of a partial-sky survey window, we also introduce an integral constraint, IC, when modelling the clustering, signal,
\begin{equation}
    w_\mathrm{gg} \to w_\mathrm{gg} + \mathrm{IC} \ .
\end{equation}
This term, which becomes important only on large scales, has the function of capturing the bias that arises from a mis-estimation of the global mean density \citep{Roche&Eales1999}. We treat this term as a nuisance parameter, such that our parameter vector reads
\begin{equation}
    \lambda = \{ b_g, A_\mathrm{IA}; \mathrm{IC} \} \ .
\end{equation}

We limit our fits to the quasi-linear regime, $r_{\rm p}>6 \mpch$, to ensure that the linear bias approximation is satisfied and the IA signal is well described by the NLA model. To perform our fits, we make use of the \textsc{Emcee} \citep{Foreman-Mackey2013emcee} package as implemented in the cosmology software \textsc{CosmoSIS}\footnote{\url{http://bitbucket.org/joezuntz/cosmosis/wiki/Home}} \citep{Zuntz2015CosmoSIS}. When analysing the chains, we exclude the first 30$\%$ of samples for a burn-in phase.

\section{Results}
\label{sec:results}

\begin{figure*}
\centering
\includegraphics[width=0.9\columnwidth]{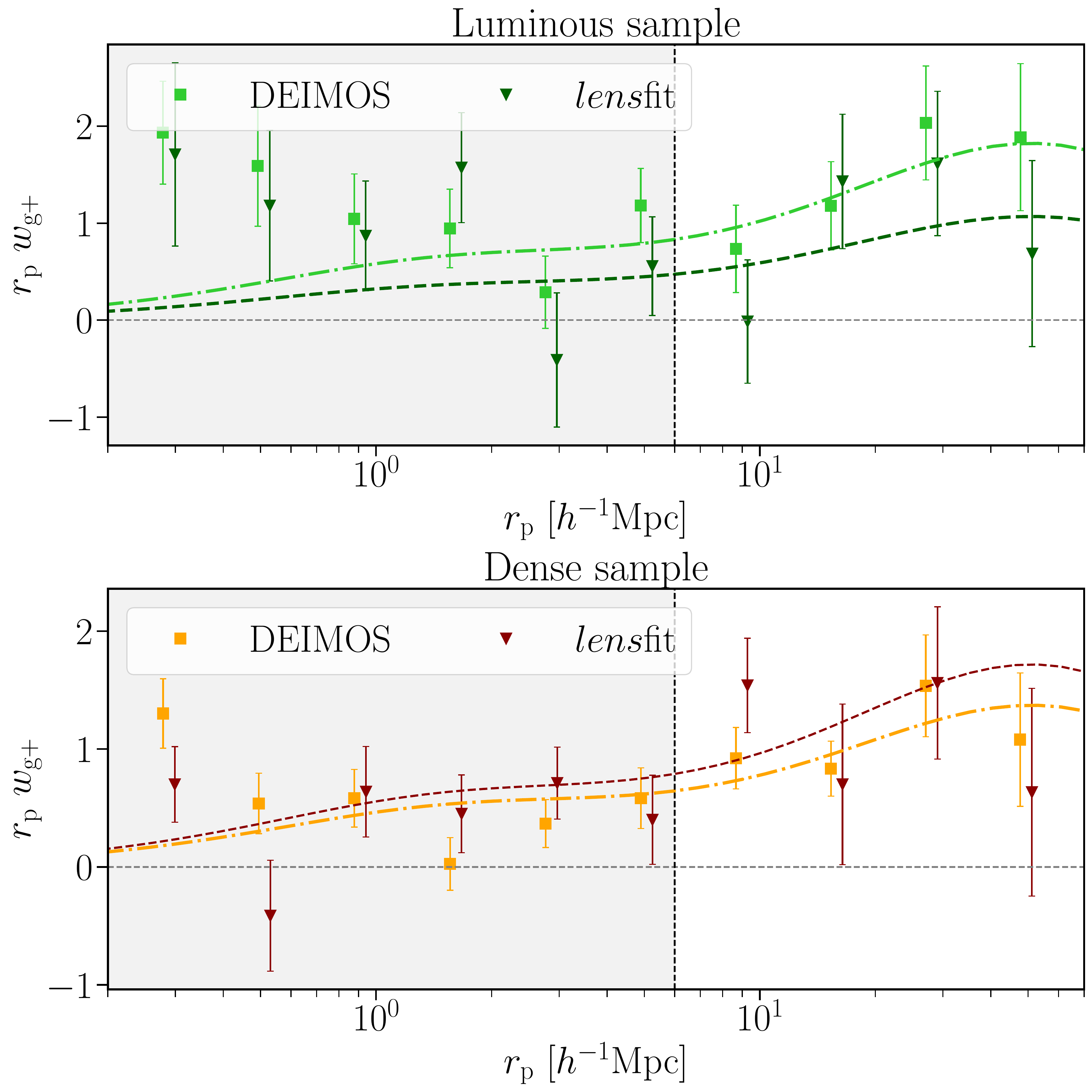}
\includegraphics[width=0.9\columnwidth]{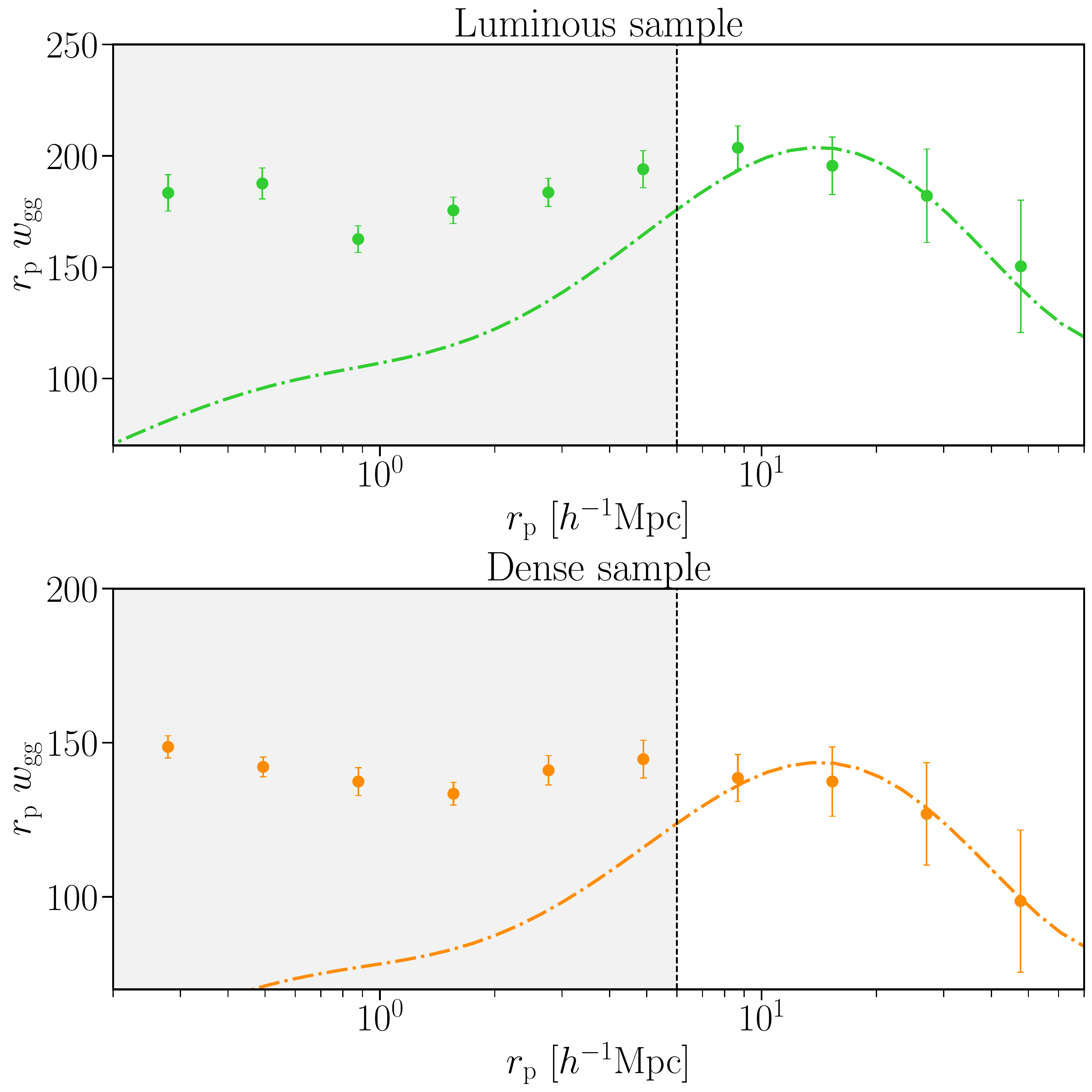}
\caption{Projected correlation functions (IA and clustering signal) measured in this work and the best-fit curve predicted by our model. \textit{Left:} Projected position-shape correlation function, $\wgp$, measured for our \lum\ (top panel) and \dense\ (bottom panel) samples. We show results for shapes measured with \textsc{DEIMOS} (light squares) and \lensfit\ (dark triangles). The best-fit models to the data with $r_{\rm p}>6 \mpch$ (indicated by the vertical dashed line), are shown as well, with the same colour scheme (\textsc{DEIMOS}: dash-dotted lines, \lensfit: dashed lines). For clarity, the \lensfit\ results have been slightly offset horizontally. \textit{Right:} Projected clustering signal, $w_\mathrm{gg}$, of the \dense\ and \lum\ samples. The dot-dashed lines corresponds to the best-fit models.  As we do not include a scale-dependent bias in our model, the mismatch between data and prediction at small scales is expected.}
\label{fig:wgp_wgg_DEIMOS_lensifit_lum_dense}
\end{figure*}

\begin{table*}
	\caption{Properties of the individual galaxy samples used in our analysis and the corresponding best-fit galaxy bias ($b_g$) and IA amplitude ($A_\mathrm{IA}$) as constrained by our model.}
	\label{tab:galinfo}
	\begin{center}
	\begin{tabular}{lcccccccr} 
		\hline
		\hline
		Samples & $\langle z \rangle$ & $N_{\mathrm{D}}$ & $N_{\mathrm{S}}$ & $[L_\mathrm{min}, L_\mathrm{max}]$ & $ \langle L \rangle/L_0$ & $b_\mathrm{g}$ & $A_\mathrm{IA}$ & $\chi^2_\mathrm{red}$ 
		\\
		\hline
		\textsc{DEIMOS} & & & & & & & & \\
		\hline 
        \dense\  & 0.44 & 173 445 & 152 832 & & 0.38 & $1.59^{+0.04}_{-0.04}$
        & $3.69^{+0.66}_{-0.65}$
        & 0.78
        \\
        \lum\  & 0.54 & 117 001 & 96 863 & & 0.64 & $2.06^{+0.04}_{-0.04}$ 
        & $4.03^{+0.81}_{-0.79}$
        & 1.19 
        \\
		\hline 
        \texttt{D1} & 0.41 & 173 445 & 39 108 & $[0.09, 1.13]$ & 0.21 & $1.60^{+0.04}_{-0.04}$ 
        & $3.02^{+1.53}_{-1.48}$
        & 1.00
        \\
        \texttt{D2} & 0.42 & 173 445 & 39 322 & [1.13, 1.43] & 0.27 & $1.60^{+0.04}_{-0.04}$
        & $1.21^{+1.63}_{-1.64}$
        & 0.91
        \\
        \texttt{D3} & 0.43 & 173 445 & 39 229 & [1.43, 1.92] & 0.35 & $1.59^{+0.04}_{-0.04}$
        & $4.11^{+1.48}_{-1.48}$
        & 1.05
        \\
        \texttt{D4} & 0.45 & 173 445 & 19 333 & [1.92, 2.81] & 0.49 & $1.59^{+0.04}_{-0.04}$
        & $3.02^{+2.37}_{-2.33}$
        & 1.52
        \\
        \texttt{D5} & 0.45 & 173 445 & 19 235 & $\geq 2.81$ & 0.89 & $1.59^{+0.04}_{-0.04}$ 
        & $8.39^{+1.04}_{-1.30}$ 
        & 0.47
        \\
        \texttt{L1} & 0.53 & 117 001 & 48 588 & $[0.29, 2.66] $ & 0.46 & $2.06^{+0.04}_{-0.04}$
        & $1.80^{+0.96}_{-0.95}$
        & 1.17
        \\
        \texttt{L2} & 0.55 & 117 001 & 24 208 & [2.66, 3.51] & 0.65 & $2.06^{+0.04}_{-0.04}$
        & $4.95^{+1.24}_{-1.21}$
        & 1.19
        \\
        \texttt{L3} & 0.56 & 117 001 & 24 067 & $\geq 3.51$ & 1.00 & $2.06^{+0.04}_{-0.04}$
        & $5.71^{+1.57}_{-1.60}$ 
        & 2.03
        \\
		\hline
		\lensfit\ & & & & & & & \\ 
		\hline
		\dense\  & 0.49 & 173 445 & 121 500 & & 0.33 & $1.60^{+0.04}_{-0.04}$
		& $4.94^{+1.24}_{-1.22}$ 
		& 1.52
		\\
        \lum\  & 0.63 & 117 001 & 84 785 & & 0.59 & $2.06^{+0.04}_{-0.04}$
        & $2.95^{+1.49}_{-1.42}$
        & 1.54
        \\
		\hline
		\textsc{DEIMOS} + \lensfit\ & & & & & & & \\ 
		\hline
        Z1 $(z \leq 0.585)$ & 0.44 & 56 754 & 56 754 & & 0.63 & $2.01^{+0.06}_{-0.06}$
        & $3.84^{+1.10}_{-1.06}$
        & 0.22 
        \\
        Z2 $(z>0.585)$ & 0.70 & 57 613 & 57 613 & & 0.61 & $2.39^{+0.08}_{-0.08}$ 
        & $3.97^{+2.02}_{-2.04}$ 
        & 2.43 
        \\
        \hline
        \end{tabular}
  \small
    \tablefoot{The galaxy properties are summarised by: the mean redshift, $\langle z \rangle$; the number of galaxies in the density (shape) sample, $N_\mathrm{D}$ ($N_\mathrm{S}$); the mean luminosity in terms of a pivot luminosity $L_0=4.6\times10^{10} h^{-2} L_{\odot}$; the bias, $b_\mathrm{g}$. To compute the ratio $\langle L \rangle / L_0$, we only consider the galaxies in the corresponding shape sample. For our $L-$cuts sub-samples, we also provide the range in luminosity they probe, $[L_\mathrm{min}, L_\mathrm{max}]$, in units of $10^{10} h^{-2} L_{\odot}$. Similarly, we provide in brackets the cut adopted to split our sample in two redshift bins. When cross-correlating different samples, $N_D$ refers to the density sample used in the correlation and the bias is the best-fit bias of the density tracer as obtained for that given measurement. All measurements are performed assuming $\Pi_\mathrm{max}=120 \mpch$. 
    Since the best-fit parameters and the medians of the marginal posterior distributions are in agreement, we quote the marginal values, while the $\chi^2$ refers to the maximum likelihood. In all cases, the degrees of freedom are 5; the $p-$values are all above 0.03, with the majority of them being in the range 0.3-0.7.}
    \end{center}
\end{table*}

The left panels in Fig.~\ref{fig:wgp_wgg_DEIMOS_lensifit_lum_dense} show the measurements of the projected position-shape correlation function $\wgp$ for the \lum\ (top panel) and \dense\ (bottom panel) samples. We present results for both the \lensfit\ (dark green triangles) and \textsc{DEIMOS} (light green squares) shape catalogues. As described in Sect.~\ref{subsec:likelihoods}, we simultaneously fit the IA and the clustering signals. We show the resulting best-fit models to measurements with $r_{\rm p}>6 \mpch$ of $\wgp$ and $w_{\rm gg}$ as solid lines in the figures. The estimates from the two shape measurement algorithms are fit independently, but given that the corresponding clustering signal is the same, here we only show the best-fit curve for the \textsc{DEIMOS} fit. The clustering measurements use the full density samples, and thus do not rely on a successful shape measurement.

We observe similar signals for the \textsc{DEIMOS} and \lensfit\ samples, with the \lensfit\ measurements having a lower S/N, because of the lack of shape measurements for galaxies with $m_r<20$. We note that we do not necessarily expect to observe the same signal, because 
\textsc{DEIMOS} contains more bright, low-redshift galaxies, whereas the
\lensfit\ sample includes fainter, distant galaxies (see Figs.~\ref{fig:z_histo_shapes} and~\ref{fig:mag_histo_dense_and_lum}). If the alignment signal depends on luminosity or redshift, the two shape samples would give different signals. In Appendix~\ref{A:DEIMOS_lensfit_comparison} we restrict the comparison to the sample of galaxies with shape measurements from both methods, and find that the average difference 
$\langle r_{\rm p}\Delta\wgp\rangle=0.003\pm0.13$ is negligible, especially compared to the amplitude of the IA signal quantified as
$\langle r_{\rm p}\wgp\rangle=0.90\pm0.17$ (\textsc{DEIMOS} shapes; see Appendix~\ref{A:DEIMOS_lensfit_comparison} for details).

We also show the models that provide the best-fit to the combined 
$w_\mathrm{gg}$ and $\wgp$ measurements in Fig.~\ref{fig:wgp_wgg_DEIMOS_lensifit_lum_dense}, and report the 
values for the bias $b_{\rm g}$ and IA amplitude $A_{\rm IA}$ in
Table~\ref{tab:galinfo}. The results for \textsc{DEIMOS} and \lensfit\  are consistent.

Our constraints on the galaxy bias of the \dense\ and \lum\ samples are in broad agreement with the values presented in \citet[][]{Vakili2020}: We find a larger bias for the \lum\ sample than for the \dense\ one, as expected by its higher luminosity and the higher redshift baseline. 

\subsection{Luminosity dependence} \label{subsec:luminosity_dependence}

\begin{figure}
\centering
\includegraphics[width=\columnwidth]{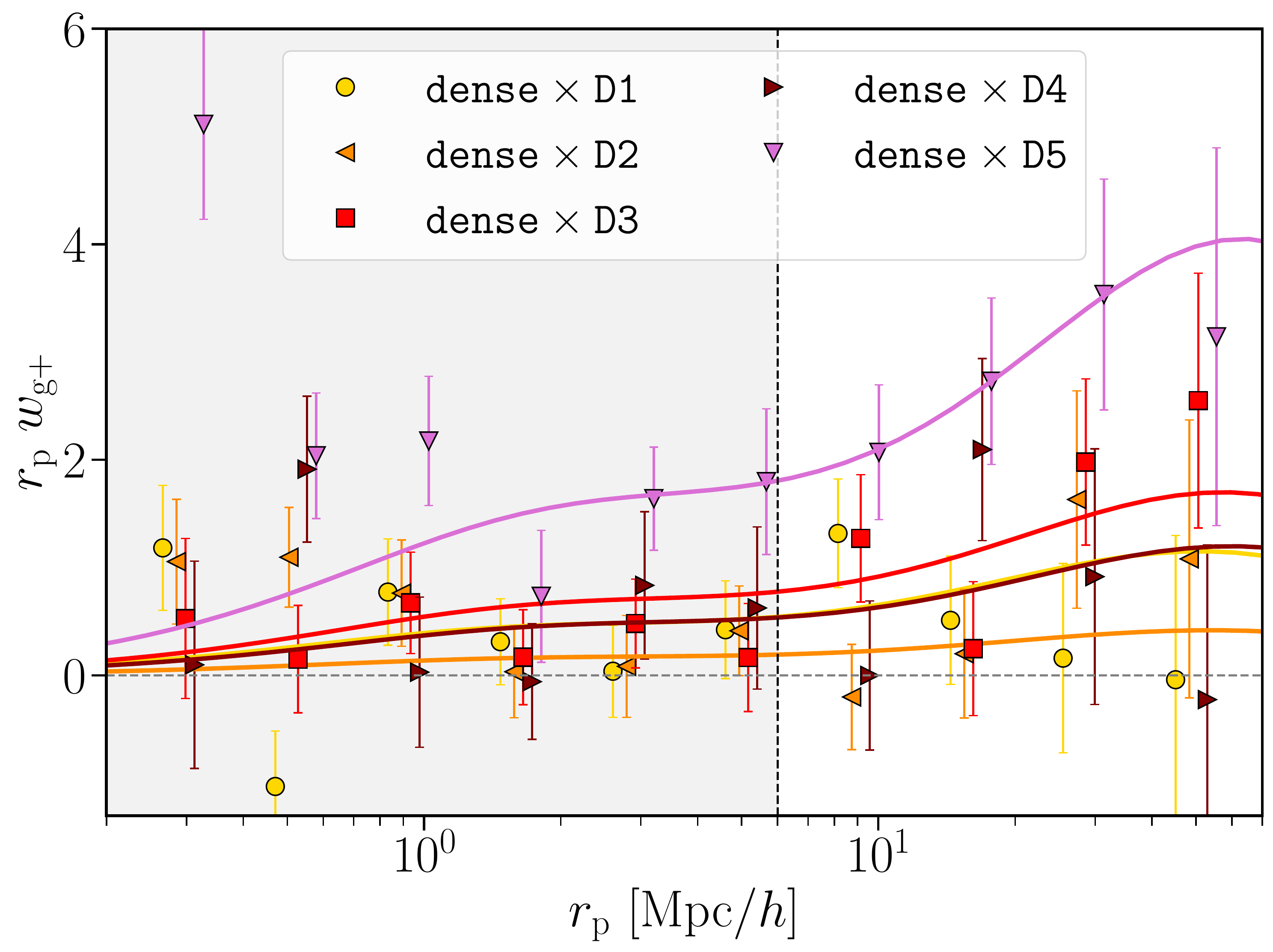}
\caption{Projected correlation function, $\wgp$, measured for the different cuts in luminosity of the \textsc{DEIMOS} \dense\ sample. The best-fit curves are plotted on top of the data points, and the fits are performed for $r_{\rm p}> 6 \mpch$. All but the yellow points have been slightly offset horizontally; to better visualise the goodness of fit, the corresponding best-fit curves have been displaced accordingly.}
\label{fig:wgp_Lcuts}
\end{figure}

Previous studies of LRGs \citep{Joachimi2011b, Singh2015} have found a significant dependence of their IA signal with luminosity, with more luminous galaxies showing stronger alignments. On average our LRG sample probes somewhat lower luminosities than those earlier studies, but the overlap with these earlier works also enables a direct comparison. Thanks to the large range in luminosity it covers, the \dense\ sample is particularly suited to explore the dependence  with luminosity. To do so, we use the \textsc{DEIMOS} shape catalogue\footnote{The internal cut at $m_r<20$ in \lensfit\ makes it less suitable for this analysis, as we have fewer galaxies at high luminosities.} and split the \dense\ LRG galaxies in five sub-samples: \texttt{D1}, \texttt{D2}, and \texttt{D3}, correspond to the lowest three quartiles in luminosity; the remaining two, \texttt{D4} and \texttt{D5}, are obtained by splitting the highest luminosity quartile into two equally sized samples. The motivation to split the quartile with the highest luminosities is that it encompasses a very large range in luminosity, which complicates the interpretation if the signal depends on luminosity (see below). Relevant details for the sub-samples are listed in Table~\ref{tab:galinfo}. We keep the \dense\ and \lum\ samples separate, in order to better isolate the effect of the luminosity dependence from any redshift evolution of the sample itself. For instance, as listed in Table \ref{tab:galinfo}, the mean redshift of the sub-samples increases somewhat from \texttt{D1} to \texttt{D5}. 

\begin{figure}
\centering
\includegraphics[width=\columnwidth]{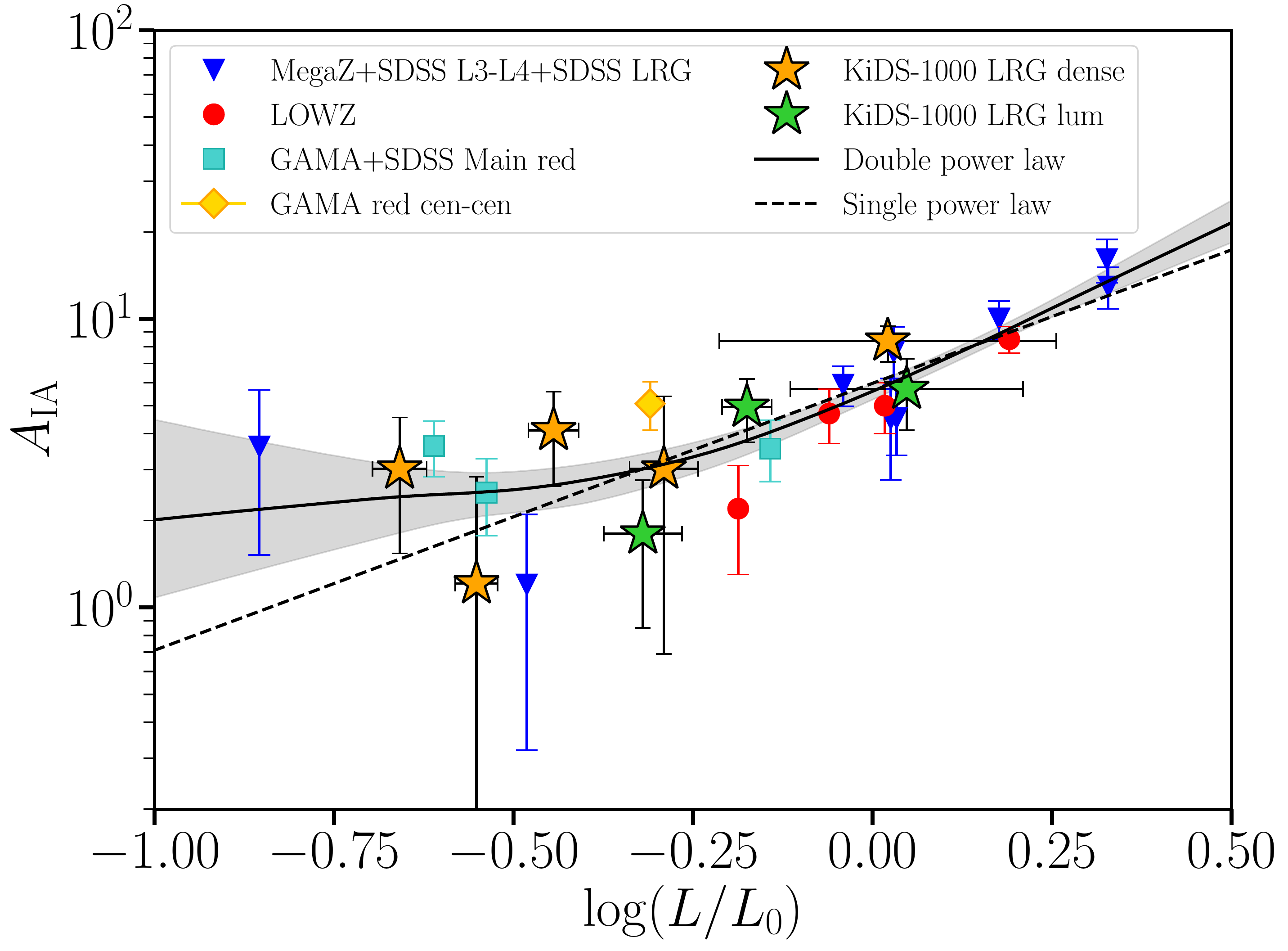}
\caption{Luminosity dependence of the IA amplitude as measured by different observational studies \citep[]{Joachimi2011b, Singh2015, Johnston2019, fortuna2020halo}; our new measurements on the LRG samples are shown as star markers. We provide horizontal error bars to indicate that the measurement is performed on a bin in luminosity, here plotted as the weighted standard deviation of the luminosity distribution of each sample, with the marker placed at the weighted mean. The solid (dashed) black line shows the median of the distribution of the MCMC sample associated with the double (single) power law; the shaded area corresponds to the $68\%$ confidence region.
}
\label{fig:A_lum}
\end{figure}

We cross-correlate the \textsc{DEIMOS} shape catalogues for the individual sub-samples with the positions of galaxies in the full \dense\ sample. In this way, we can disentangle the luminosity dependence of the IA signal from the luminosity dependence of the density tracer (brighter galaxies are typically found in denser environments). The measurements and the best-fit models are presented in Fig.~\ref{fig:wgp_Lcuts}. In Table~\ref{tab:galinfo} we list the best-fit values for the galaxy bias $b_{\rm g}$ and IA amplitude $A_{\rm IA}$, as well as the reduced $\chi^2$, as before, using the measurements for $r_{\rm p}>6 \mpch$. We also show the measurements in Fig.~\ref{fig:A_lum} as orange stars as a function of $L/L_0$, where $L_0=4.6 \times 10^{10}h^{-2} L_{\odot}$. 

\begin{figure}
\centering
\includegraphics[width=\columnwidth]{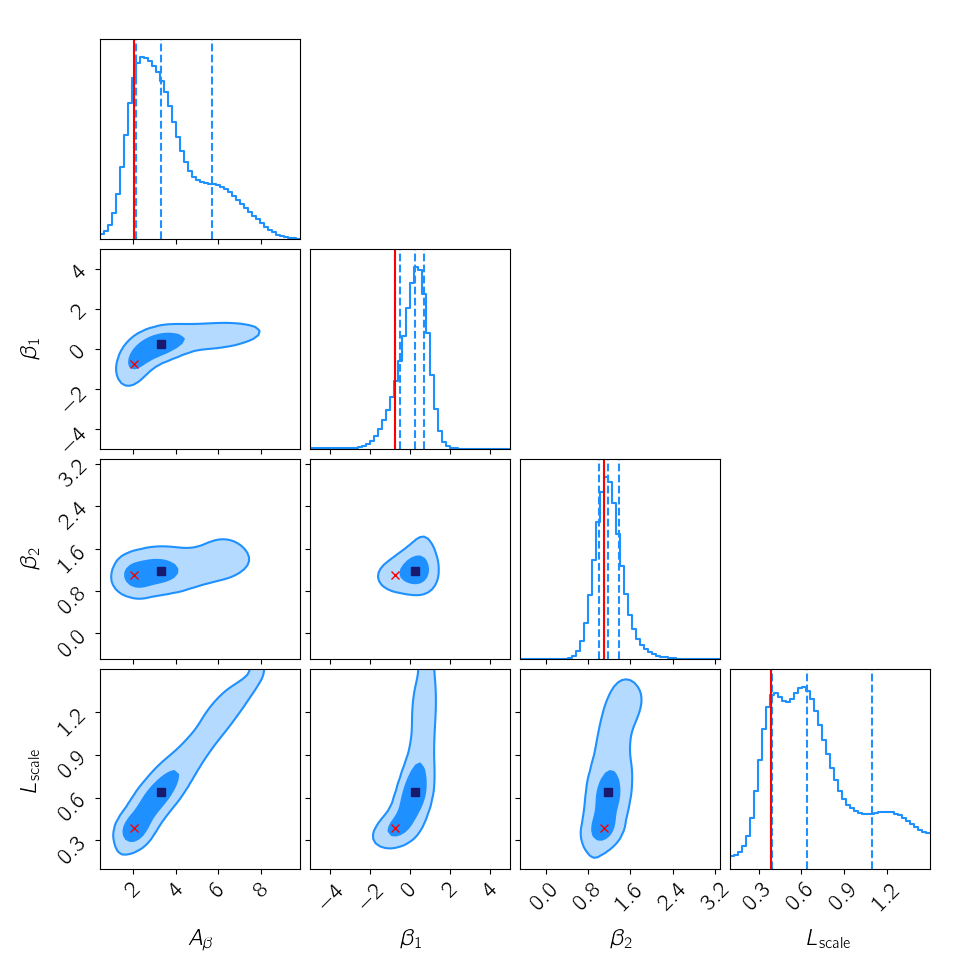}
\caption{Constraints on the double power law parameters described in equation ~\ref{eq:double_powerlaw} by jointly fitting all the measurements in Fig.~\ref{fig:A_lum}. The red crosses indicate the value of the parameters that maximise the likelihood, while the blue squares correspond to the medians.}
\label{fig:a_lum_mcmc}
\end{figure}

\begin{table*}
	\caption{Best-fit parameters of the single and double power law fit on the measurements in Fig.~\ref{fig:A_lum}.}
	\label{tab:all_lum_fit_pars}
	\begin{center}
	\begin{tabular}{lcccccr} 
		\hline
		\hline
		Model & $A_{\beta}$ & $\beta_1$ & $\beta_2$ & $L_{\rm break}$ & $\chi^2$/dof & dof\\
		\hline
		Double power law & $3.28^{+2.41}_{-1.17}$ (2.01) & $0.26^{+0.42}_{-0.77}$ (-0.75) & $1.17^{+0.21}_{-0.17}$ (1.11) & $0.64^{+0.45}_{-0.24} L_0$ ($0.39 L_0$) & 1.36 (1.33) & 22\\
        Single power law & $5.98^{+0.27}_{-0.27}$ (6.0) & $0.93^{+0.11}_{-0.10}$ (0.92) & - & $L_0$ & 1.61 (1.61) & 24\\
        \hline
    \end{tabular}
    \small
    \tablefoot{The listed values correspond to the medians of the marginal posterior distributions, and the associated errors correspond to the 16th and 84th percentiles, while in brackets we report the parameters that maximise the likelihood. The same scheme is adopted for the corresponding reduced $\chi^2$. $L_{\rm break}$ is the pivot luminosity that enters in the denominator of the power law argument. For convenience, the slope of the single power law model is here reported as $\beta_1$.}
    \end{center}
\end{table*}

We repeat the same analysis for the \lum\ sample, which we divide in three bins, with a similar bin refining approach as for the \dense\ sample (in this case \texttt{L1} contains half of the \lum\ galaxies, while \texttt{L2} and \texttt{L3} the remaining quarters). The best-fit amplitudes for these samples are reported in Table~\ref{tab:galinfo}, and presented as green stars in Fig.~\ref{fig:A_lum}. In the luminosity range where the \lum\ and \dense\ samples overlap, we find the results between the two samples to be compatible. The \lum\ sample seems to show a more pronounced luminosity dependence compared to the \dense\ sample, which can either be an effect of being brighter overall (from \texttt{L1} to \texttt{L3}, $L/L_0 = 0.46, 0.64, 1.01$) or 
due to the satellite fraction being lower (see Sect. \ref{subsec:satellite_fraction}), or a combination of the two. We note that the measurements of the \texttt{L3} sample appear to scatter more than the covariance predicts, which results in higher $\chi^2$. 
A similar issue is present in the \texttt{D4} sample and it is visible in Fig.~\ref{fig:wgp_Lcuts}.

The horizontal error bars in Fig.~\ref{fig:A_lum} indicate the weighted standard deviation of the luminosity distribution within the bin for each sample, with the measurement placed at the luminosity-weighted mean of the bin. If the range is too large, and the IA signal varies within the bin, the resulting amplitude is difficult to interpret, and may even appear discrepant. For instance, when we combine the \texttt{D4} and \texttt{D5} samples we obtain $A_\mathrm{IA} = 6.70^{+1.15}_{-1.14}$. We note, however, that the luminosity range probed by this combined bin is particularly extended, and the high signal measured is mainly driven by the galaxies in the high luminosity tail of the bin (\texttt{D5}, $A_\mathrm{IA} = 8.39^{+1.04}_{-1.30}$). The other half of the bin has a relatively low signal with very large uncertainties (\texttt{D4}, $A_\mathrm{IA}=3.02^{+2.37}_{-2.33}$). This is relevant because it suggests that the alignment of galaxies with luminosities below $L/L_0 \sim 0.60-0.70$ hardly depends on luminosity, and thus with a similar amplitude to \texttt{D1} and \texttt{D3}, the smaller sample is less constraining. As soon as we exceed this approximate threshold, the signal increases significantly, suggesting a luminosity dependence. This overall picture is enhanced when we also consider previous results for LRGs \citep[][]{Joachimi2011b, Singh2015, Johnston2019, fortuna2020halo}\footnote{The GAMA points \citep{Johnston2019} have been adjusted to homogenise the units convention, as discussed in \citet[][]{fortuna2020halo}.}.
These are also shown in Fig.~\ref{fig:A_lum}. We investigate how well the current measurements support the picture of a single or double power law by fitting the data points in Fig.~\ref{fig:A_lum}, assuming them to be uncorrelated. For each data point, we only use the quoted $L/L_0$ as we do not have the underlying luminosity distribution for most of the measurements. We propose a double power law with knee at $L_{\rm break}$, amplitude $A_{\beta}$ and slopes $\beta_{1,2}$:
\begin{equation}\label{eq:double_powerlaw}
    A(L) = A_{\beta} \left( \frac{L}{L_{\rm break}} \right)^{\beta} \rm {with} \ \begin{cases} \beta = \beta_1 & {\rm for} \ L< L_{\rm break}\\
    \beta = \beta_2 & {\rm for} \ L> L_{\rm break}
    \end{cases}
\end{equation}
and fit for 
\begin{equation}
    \lambda = \left\{ A_{\beta}, \beta_1, \beta_2, L_{\rm scale} \right\} \ ,
\end{equation}
where $L_{\rm scale} = L_{\rm break}/L_0$. We explored the parameter space using a MCMC and assuming a Gaussian likelihood. Figure~\ref{fig:a_lum_mcmc} shows our parameter constraints, while the model prediction is shown in Fig~\ref{fig:A_lum} as a solid black line. Our best-fit parameters are reported in Table ~\ref{tab:all_lum_fit_pars}\footnote{We note that the parameters that maximise the likelihood differ from the medians of the posterior distributions as a consequence of the degeneracies between the parameters. This is particularly evident for $\beta_1$, which has negative slope, $\beta_1 = -0.75$.}. We repeated the same analysis assuming a single power law, as parametrised in \citet{Joachimi2011b}. The best-fit parameters are also reported in Table~\ref{tab:all_lum_fit_pars}.
The larger $\chi^2/$dof of the single power law compared to the double power law suggests that the latter is a better description of our current data, although the scatter between the points at low $L$ is still too large to draw definitive conclusions and the data are also mildly inconsistent in that regime. The degeneracy between the parameters, and in particular between $A_{\beta}$ and $L_{\rm scale}$, shows that the data can weakly constrain the model. Nevertheless, the emerging picture seems to support more the broken power law scenario presented in \citet[][]{fortuna2020halo}, but with a transition luminosity around $0.4-0.6 L_0$, also in line with the results from simulations by \citet{Samuroff2020}. The double power law is also supported by the fact that the alignment of redMaPPer clusters \citep[][]{vanUitert2017,Piras2018}, not included in this analysis, forms a smooth extension towards higher mass of the alignment observed for the high luminosity LRGs. This result is hard to reconcile with a single shallow power law, but finds a natural framework in the double power law scenario, where the slope of the relation at high luminosities recovers the trend in \citet[][]{Joachimi2011b, Singh2015}. 

We caution that this analysis does not aim to be fully comprehensive, but rather to provide a sense of the current trends. A proper analysis should jointly fit all of the measurements incorporating the full luminosity distributions of each sample, as well as accounting for the presence of satellites, which might suppress the signal at low luminosities. 

\subsection{Redshift dependence} \label{subsec:redshift_dependence}

\begin{figure}
\centering
\includegraphics[width=\columnwidth]{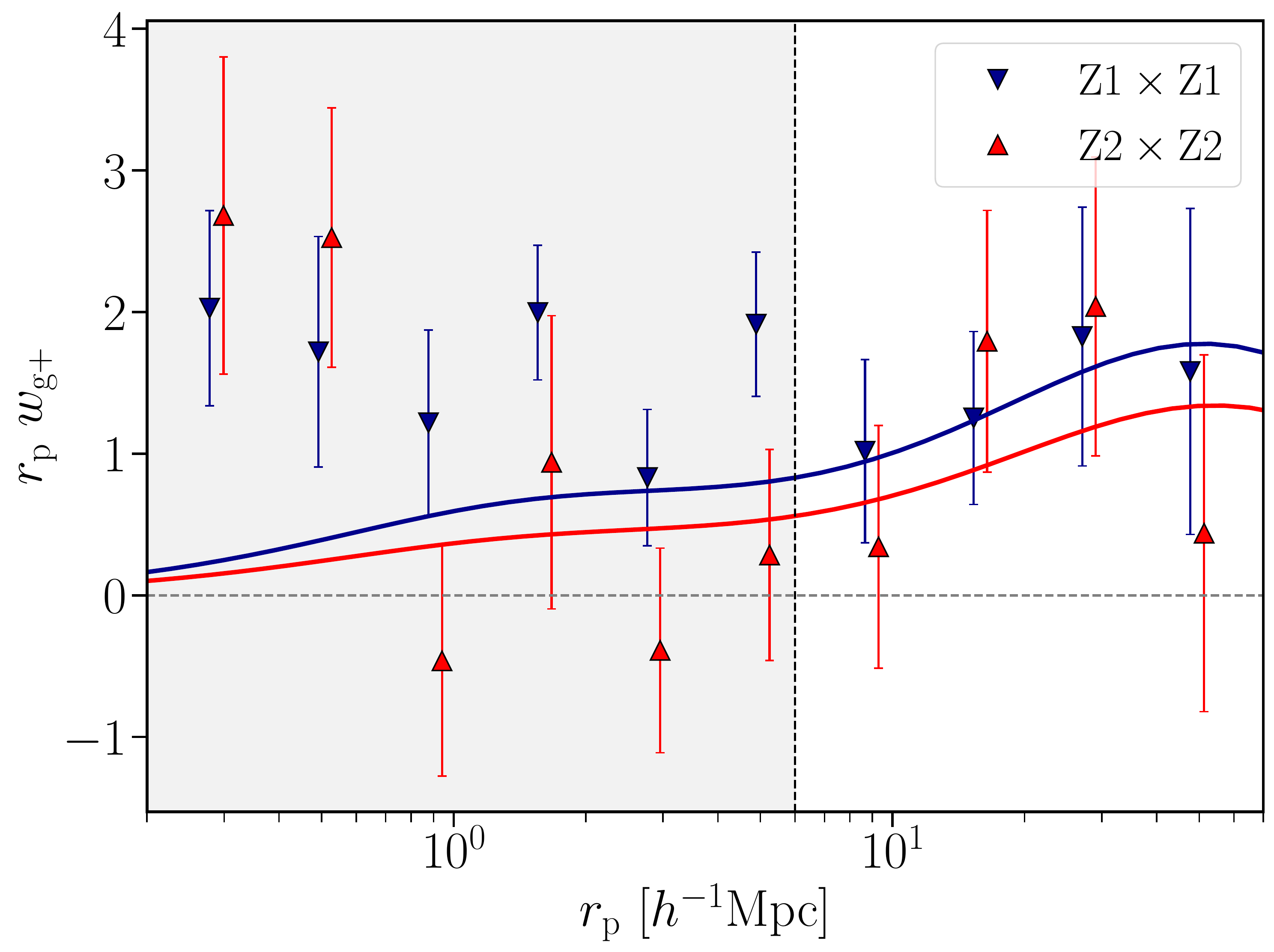}
\caption{Projected correlation function, $\wgp$, measured on our different cuts in redshift of the \lum\ sample. The best-fit curves are plotted on top of the data points, and the fits are performed for $r_{\rm p}> 6 \mpch$. The red points are slightly displaced for clarity and the corresponding best-fit curve has been displaced accordingly.}
\label{fig:wgp_zcuts}
\end{figure}

Having assessed that the two shape measurements produce compatible IA signals and that their calibrations are robust, we merge
the two shape catalogues to span the largest possible range in redshift. This allows us to extend the sample from the low-$z$, high S/N galaxies, where only \textsc{DEIMOS} provides shapes, to the high-$z$, low S/N galaxies, where we preferentially measure the shapes via \lensfit. In the case of overlap between \textsc{DEIMOS} and \lensfit, we select the \textsc{DEIMOS} shapes. We only focus on the \lum\ sample as we are interested in a long redshift baseline with the same luminosity cut. In this way, we can probe the redshift evolution of the sample, without confusing the results with any luminosity dependence.

Our final catalogue contains 115\,322 galaxies that we split at $z=0.585$, which roughly provides two equally populated bins. We call these two samples Z1 and Z2. The measurements for 
$\wgp$ are presented in Fig.~\ref{fig:wgp_zcuts}. The best-fit values for the two redshift bins are listed in Table~\ref{tab:galinfo} and agree within their error bars, despite their mean redshift being $\langle z \rangle = 0.44$ and $\langle z \rangle = 0.70$, respectively. 

We note that the $\chi^2$ of our Z2 sample is quite high: This is driven by the poor fit of the clustering signal. We attribute this to our photo$-z$, which at high redshift are less reliable. We note, however, that the uncertainty in the IA amplitude is large enough to absorb the inaccuracies in $p(z_{\rm spec}|z_{\rm phot})$, such that modifying the redshift distributions has little impact on the recovered IA amplitude.

\begin{figure}
\centering
\includegraphics[width=\columnwidth]{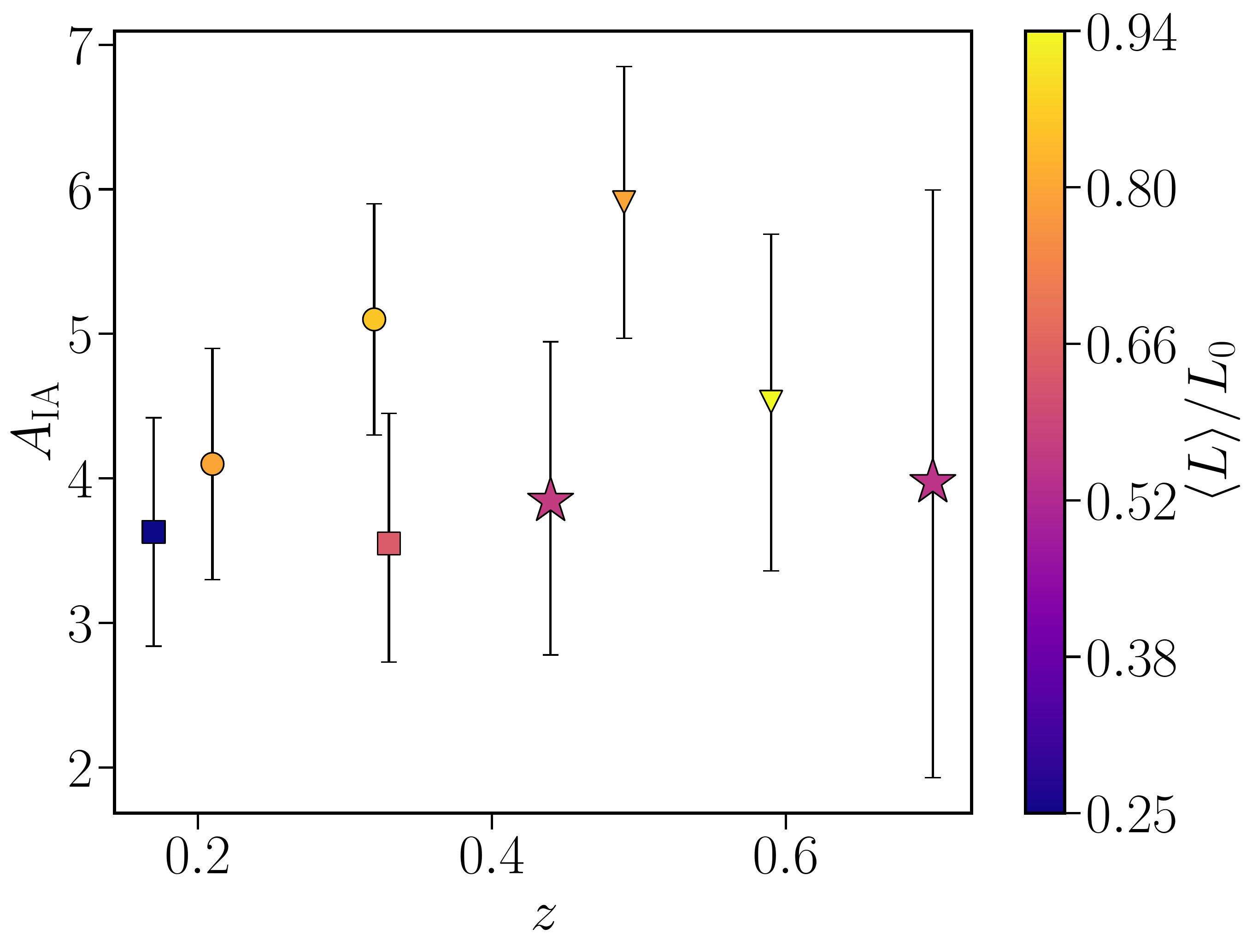}
\caption{Intrinsic alignment amplitude, $A_\mathrm{IA}$, as a function of redshift and luminosity for different best-fit values in the literature. Different markers refer to different studies and are colour-coded based on their luminosity: MegaZ \citep[][]{Joachimi2011b} is shown as triangles, LOWZ \citep{Singh2015} as circles, GAMA \citep[][]{Johnston2019} as squares and the LRG \lum\ sample Z1 and Z2 as stars.}
\label{fig:A_z_all}
\end{figure}

Figure~\ref{fig:A_z_all} compares our results with the best-fit amplitudes at various redshifts found by previous studies \citep{Joachimi2011b,Singh2015,Johnston2019}. The colour of the data points reflects the luminosity of the sample used to measure the signal\footnote{The colour of the marker corresponds to the bin centre, which may not be sufficient if the range in luminosity is large, as it
is typically the case for these samples. The information provided by the colour has therefore only qualitative meaning and should be considered as such.}. As previously discussed, galaxies with different luminosities may manifest different levels of IA, and hence even with a lack of  redshift dependence, we should still expect points at different amplitudes: the bottom part of the plot should be mainly populated by darker points and the upper part by brighter points. Figure ~\ref{fig:A_z_all} confirms this scenario: overall, the points exhibit a similar alignment and the scatter between the different points is consistent with the extra luminosity dependence. We can conclude that there is little evidence for a strong redshift dependence of the IA signal.

\section{Conclusions}
\label{sec:conclusions}

We have constrained the IA signal of a sample of LRGs selected by 
\cite{Vakili2020} from KiDS-1000, which images $\sim 1000$ deg$^2$.
These data allowed us to investigate the luminosity dependence and the redshift evolution of the signal. To do so, we measured the shapes of the LRGs with two different algorithms, \textsc{DEIMOS} and \lensfit. We used custom
image simulations to calibrate and correct the residual biases that arise from measurements of noisy images. 

We used the calibrated ellipticities to compute the projected position-shape correlation function $\wgp$ and analyse the signals obtained by the two different algorithms independently, thus exploring the dependence of IA on the specific shape method employed. We found \lensfit\ measurements to be overall noisier than the \textsc{DEIMOS} ones and we attributed this to the prevalence of faint galaxies in the sample, due to the internal magnitude cut in the \lensfit\ algorithm. Because bright galaxies typically carry more alignment signal, this cut, which removes galaxies with $m_r<20$, can potentially reduce the IA contamination in KiDS cosmic shear analyses, which employ \lensfit\ as the shape method. For a sub-sample of galaxies, where both shape methods return successful measurements of the shapes, we find a remarkable agreement in the measured $\wgp$, with a difference in the signal of $0.003 \pm 0.13$ (amplitude of a fitted power law).

We explored the luminosity dependence and the redshift evolution independently, selecting our galaxies in such a way that ensures the two do not mix.  Within the luminosity range probed by the measurements our results agree with previous studies \citep[][]{Joachimi2011b, Singh2015, Johnston2019}. However, a single power law fit, as was used in \cite{Joachimi2011b} and \cite{Singh2015} does not describe the measurements well. Instead, our results suggest a more complex dependence with luminosity: for $L_r\lesssim 2.9\times 10^{10}h^{-2}L_{r,\odot}$
the IA amplitude does not vary significantly, whereas the signal rises rapidly at higher luminosity. This also has implications for the width of the luminosity binning, as the use of broad bins may complicate the interpretation of the measurements. Analyses that aim to combine these measurements to model the  luminosity dependence should incorporate the underlying luminosity distributions to properly link the signal to the galaxy luminosity. Nevertheless, we provide a preliminary fit on the current measurements available in the literature and found that the data are best described by a broken power law. This result can already be used by cosmic shear analyses to improved their modelling of the IA carried by the red galaxy population. We remind the reader that this sample is not representative of the galaxy population. Different galaxy samples carry different alignment signals and should thus be individually modelled as described in \citet{fortuna2020halo}.

To probe the redshift dependence of the IA signal with the largest baseline to date, we merge the \textsc{DEIMOS} and \lensfit\ catalogues. We find no evidence for redshift evolution of the IA signal. This result is in line with previous studies of LRG samples \citep[][]{Joachimi2011b, Singh2015}, and it is consistent with the current paradigm that IA is set at the moment of galaxy formation. However, it is also possible that galaxy mergers counteract the evolution of the tidal alignment, such that the net signal does not change. Further improvements in the measurements are needed to distinguish between scenarios.

\section*{Acknowledgements}

We thank Sandra Unruh for providing useful comments to the manuscript. MCF, AK, MV and HH acknowledge support from Vici grant 639.043.512, financed by the Netherlands Organisation for Scientific Research (NWO). HH also acknowledges funding from the EU Horizon 2020 research and innovation programme under grant agreement 776247. HJ acknowledges support from the Delta ITP consortium, a program of the Netherlands Organisation for Scientific Research (NWO) that is funded by the Dutch Ministry of Education, Culture and Science (OCW). We also acknowledge support from: the European Research Council under grant agreement No. 770935 (AHW, HHi) and No. 647112 (CH and MA); the Polish Ministry of Science and Higher Education through grant DIR/WK/2018/12 and the Polish National Science Center through grants no. 2018/30/E/ST9/00698 and 2018/31/G/ST9/03388 (MB);  the Max Planck Society and the Alexander von Humboldt Foundation in the framework of the Max Planck-Humboldt Research Award endowed by the Federal Ministry of Education and Research (CH); the Heisenberg grant of the Deutsche Forschungsgemeinschaft Hi 1495/5-1 (Hi); the Royal Society and Imperial College (KK) and from the Science and Technology Facilities Council (MvWK).

The MICE simulations have been developed at the MareNostrum supercomputer (BSC-CNS) thanks to grants AECT-2006-2-0011 through AECT-2015-1-0013. Data products have been stored at the Port d'Informaci$\acute{\mathrm{o}}$ Cient$\acute{\mathrm{i}}$fica (PIC), and distributed through the CosmoHub webportal (cosmohub.pic.es). Funding for this project was partially provided by the Spanish Ministerio de Ciencia e Innovacion (MICINN), projects 200850I176, AYA2009-13936, AYA2012-39620, AYA2013-44327, ESP2013-48274, ESP2014-58384 , Consolider-Ingenio CSD2007- 00060, research project 2009-SGR-1398 from Generalitat de Catalunya, and the Ramon y Cajal MICINN program.

Based on data products from observations made with ESO Telescopes at the La Silla Paranal Observatory under programme IDs 177.A- 3016, 177.A-3017 and 177.A-3018, and on data products produced by Tar- get/OmegaCEN, INAF-OACN, INAF-OAPD and the KiDS production team, on behalf of the KiDS consortium. OmegaCEN and the KiDS production team acknowledge support by NOVA and NWO-M grants. Members of INAF-OAPD and INAF-OACN also acknowledge the support from the Department of Physics $\&$ Astronomy of the University of Padova, and of the Department of Physics of Univ. Federico II (Naples). 

\textit{Author contributions:} All authors contributed to the development and writing of this paper. The authorship list is given in three groups: the lead authors (MCF, HH, HJ, MV, AK, CG) followed by two alphabetical groups. The first alphabetical group includes those who are key contributors to both the scientific analysis and the data products. The second group covers those who have either made a significant contribution to the data products, or to the scientific analysis.



\bibliographystyle{aa}
\bibliography{lrgbiblio}



\appendix

\section{m-bias calibration} \label{A:mbias_calibration}

In this Appendix, we detail our procedure to calibrate the $m$-bias in our shape measurements. We follow the same procedure for both \textsc{DEIMOS} and \lensfit, but we present the results separately.

\subsection{\textsc{DEIMOS}} \label{AA:DEIMOS_calibration}
One of the key features of \textsc{DEIMOS} that was exploited by \cite{Georgiou2019b} is that the weight function that is used to measure the moments of the surface brightness distribution can be adjusted. As explained in Sect.~\ref{sec:shape_measurements}, we follow \cite{Georgiou2019} and adopt a Gaussian weight function with a width $r_{\rm iso}$. However, not only the radial profile can be changed, but one can also choose between a circular or an elliptical weight function. Hence, before proceeding with the shape calibration, we investigate which choice of weight function would suit our data best. 

In both cases, the weight function is centred on the centroid of the galaxy, with the size and ellipticity iteratively matched to those measured for the galaxy \citep[see][for details]{Georgiou2019}. While an elliptical weight function matches the shape of an elliptical galaxy better, a circular one generally performs better on small and faint objects.

The circular weight function performs similar to the elliptical weight function for low-to-intermediate S/N (S/N$<60$), but with
an overall constant bias of $\sim0.2$ as the S/N increases. Hence, the elliptical weight function performs significantly better for more than half of the (real) galaxy sample, which motivates our choice to adopt an elliptical weight function in our analysis.

\textsc{DEIMOS} measured the shapes of 13\,301 simulated LRGs from our image simulations, and we use these to calibrate our ellipticity estimates. To do so, we first explore the dependence of the $m$-bias on the individual galaxy parameters S/N and $R$, as discussed in Sect. \ref{sec:shape_measurements}. Figure~\ref{fig:mbias_gals_par} indicates
that $m(R)$ is well described by a polynomial curve, which we truncate at degree 3, $p(R)= p_1 R + p_2 R^2 + p_3 R^3$, while $c(\mathrm{S/N})$ is well described by: $d(\mathrm{S/N}) = d_1/\sqrt{\mathrm{S/N}} + d_2/\mathrm{S/N}$. 

We have tested different combinations of the two functions $m(\mathrm{S/N})$ and $m(R)$, and explored if higher-order polynomials are needed: while the fit to $m(R)$ is indeed better described by a polynomial of degree 5, we stress that we are not interested to reproduce all of the noisy features in the data, but rather to capture the trend in the two components. We therefore keep the number of the parameters as low as possible. This is also motivated by the fact that the image simulations suffer from galaxy repetitions. 

The final expression for our empirical correction for the \textsc{DEIMOS} measurements is then:
\begin{equation} \label{eq:mbias_surface}
    m(\mathrm{S/N}, R) = b_0 +  \frac{1+d(\mathrm{S/N})}{1+p(R)} \ .
\end{equation}

To find the best-fit parameters in \ref{eq:mbias_surface}, we re-compute the value of the $m-$bias by binning the data in 64 regions using the $k-$means algorithm\footnote{\url{https://scikit-learn.org/stable/modules/generated/sklearn.cluster.KMeans.html}}. We then measure the bias for the two components $\epsilon_{1,2}$ in each region, identifying the bin coordinate in S/N and $R$ as their mean value within the bin. We then fit the average of the two components $(m_1+m_2)/2$ with equation \ref{eq:mbias_surface}. Some of the galaxies have very small shape measurement errors, and to avoid them dominating the fit, we also added an intrinsic scatter $\sigma_\mathrm{int}$ to our error-bars. This accounts for the fact that the number of unique galaxies in our simulations is limited and mitigates the importance of the highly resolved ones. The intrinsic scatter $\sigma_\mathrm{int}$ is chosen such that the reduced $\chi^2$ is $\sim 1$. The best-fit parameters are reported in Table~\ref{tab:mbias_surface_parameters}. We stress here that since we are only correlating shapes with positions, we are not interested in a perfect calibration of the bias per galaxy, but rather want to ensure that the mean ellipticity of an ensemble is unbiased.

\begin{table}
	\caption{Best-fit parameters for the empirical correction of the two-dimensional multiplicative bias surface (Sect. \ref{subsec:image_simulations}).}
	\label{tab:mbias_surface_parameters}
	\begin{tabular}{lcc}
		\hline
		\hline
		Parameter & \textsc{DEIMOS} & \lensfit \\
		\hline
		$b_0$ & $-0.895$ & $0.1794$ \\
		$d_1$ & $5.238$ & $-5.081$ \\
		$d_2$ & $-0.006$ & $1.292$ \\
		$p_1$ & $-1.900$ & $-0.972$ \\
		$p_2$ & $5.147$ & $0.669$ \\
		$p_3$ & $-3.148$ & $0.783$ \\
		$p_4$ & -- & $-0.698$ \\
		\hline
		\hline
	\end{tabular}
\end{table}

\subsection{\lensfit} \label{AA:lensfit_calibration}

In the case of \lensfit\ we follow a very similar procedure to calibrate the residual $m$-bias. \lensfit\ successfully measured the shapes of 17\,573 simulated galaxies, which are used for the calibration. The dependence of the $m$-bias with S/N can be described by the same parametrisation that we used for the \textsc{DEIMOS} sample, $d(\mathrm{S/N})$, while $m(R)$ is better described by a polynomial of degree four, $p(R)=p_1 R + p_2 R^2 + p_3 R^3 + p_4 R^4$.

The combination that best reproduces our measurements of the $m$-bias in k-means cells of the two-dimensional space (S/N,$R$) is
\begin{equation}
    m(\mathrm{S/N},R) = b_0 + d(\mathrm{S/N}) + p(R) \ ,
\end{equation}
with the specific values of the parameters reported in Table~\ref{tab:mbias_surface_parameters}. We note that compared to \textsc{DEIMOS}, the \lensfit-bias is small, and hence so is our correction.

\section{Redshift distributions} \label{A:redshift_distributions}

\begin{figure}
\centering
\includegraphics[width=\columnwidth]{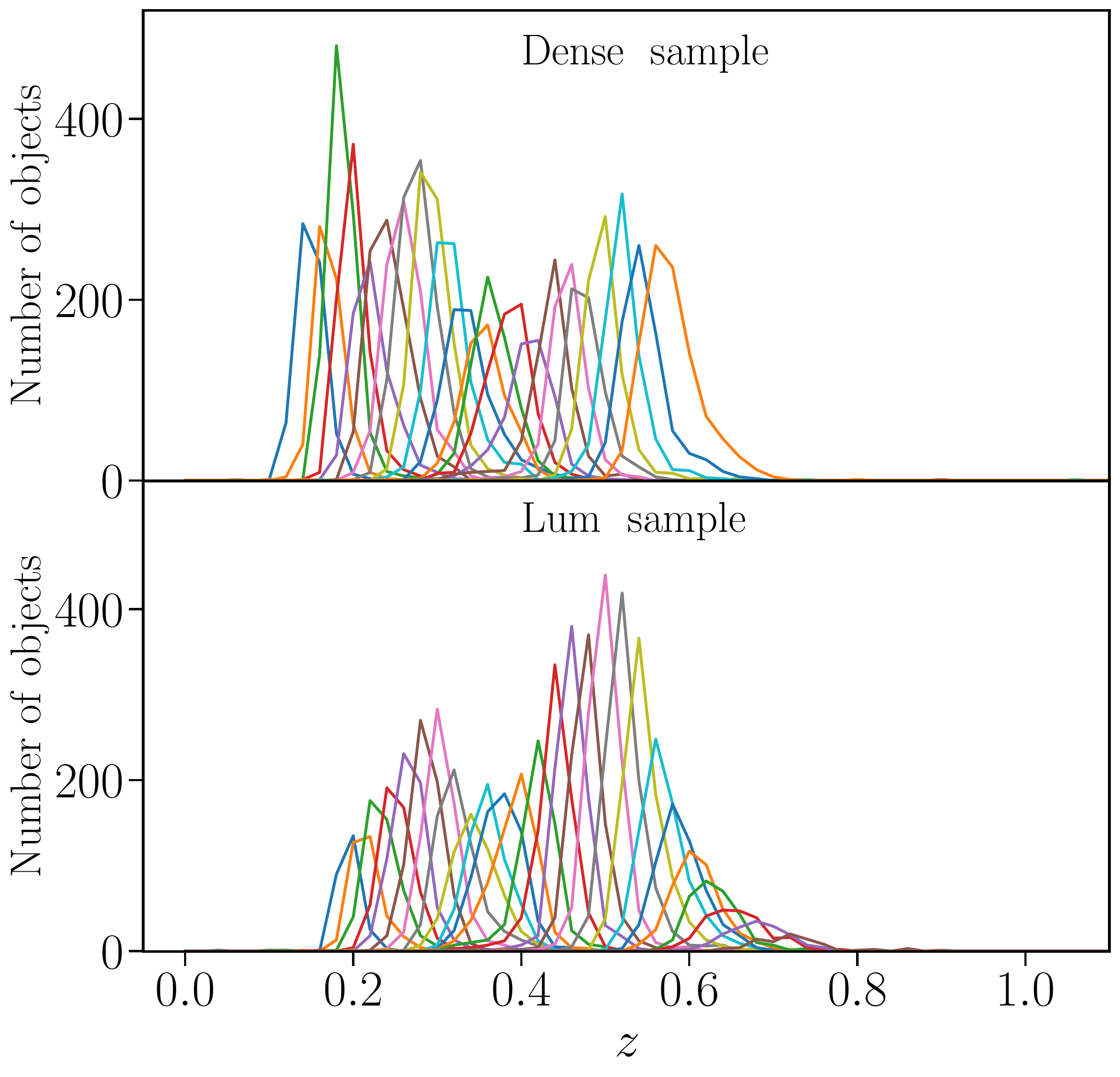}
\caption{The $p(z_\mathrm{spec}|z_\mathrm{phot})$ of our \dense\ and \lum\ samples.}
\label{fig:pofz_dense_and_lum}
\end{figure}

We describe here the redshift distributions, $p(z_{\rm spec}|z_{\rm phot})$, employed in our analysis as reported in Sect.~\ref{sec:theoretical_framework} and which are used in the computation of the angular power spectra in Eq.~(\ref{eq:C_n_eps}). We bin the galaxies for which we have spectroscopic redshifts in bins of $\Delta z_{\rm phot}^{\dense} = 0.0146(1+z)$ and $\Delta z_{\rm phot}^{\lum} = 0.0139(1+z)$ with an iterative procedure; this constructs unequal binning whose size increases with $z$. The last bin is adjusted to avoid spurious results: if the maximum redshift found with the iterative procedure exceed the maximum redshift of the sample, we remove the last bin and extend the second-to-last up to $z_{\rm max}$.  In the case of the \lum\ sample we further increase the scatter at high redshift to account for the increasing uncertainty of our photometric redshifts: for $z>0.7$ we increase the bin width to $\Delta z_{\rm phot}^{\lum} = 0.027$. We adopt the same approach for the Z1 and Z2 samples, for which we use, respectively, $\sigma_z = 0.0133$ and $0.0190$. We use the resulting spec-$z$ histograms in our analysis.  
We employ the same conditional redshift distributions for both our density and shape samples; while this is a very good approximation for \textsc{DEIMOS}, \lensfit\ lacks bright galaxies that would populate our spec-$z$, and thus this approximation might partially be responsible for the worse fit of the model. 

We have tested that our IA constraints are only marginally dependent on the width of the bins adopted, and the changes in the best-fit amplitude are subdominant to the statistical uncertainty.

\section{Contamination from galaxy-galaxy lensing} \label{A:lensing_contamination}

\begin{figure}
\centering
\includegraphics[width=\columnwidth]{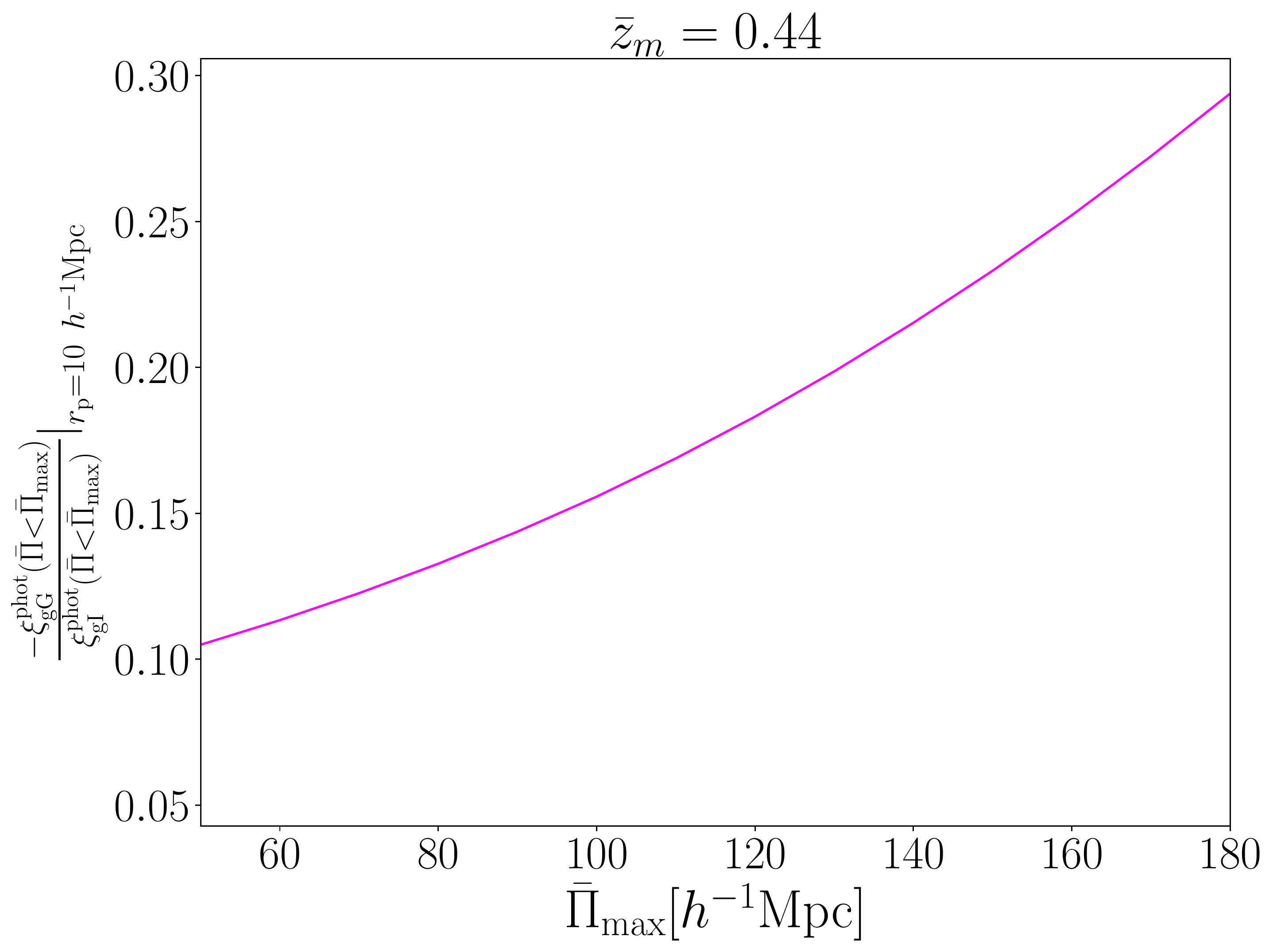}
\caption{Ratio of the cumulative galaxy-galaxy lensing signal over the cumulative IA signal as a function of $\Pi_{\rm max}$ at the mean redshift of the \dense\ sample, $z=0.44$. }
\label{fig:gG_gI_ratio}
\end{figure}

As discussed in Sect. \ref{sec:contaminations}, galaxy-galaxy lensing is the main astrophysical contaminant to our signal. Here, we focus on its dependence on the line-of-sight integration range. The lensing and the IA signals scale differently with distance: this can be used to maximise the signal and avoid an excess of contamination. In this Appendix we therefore explore in more detail the modelling of the galaxy-galaxy lensing and how this has guided our choice for the value of $\Pi_\mathrm{max}$.

Figure~\ref{fig:gG_gI_ratio} shows the amount of lensing contamination as a function of the maximum $\Pi$ used in the integral along the line-of-sight. We illustrate it by plotting the cumulative contribution of the galaxy-galaxy lensing over the one of IA for different values of the truncation, $\Pi_{\rm max}$. To generate the signal, we use the $p(z_{\rm spec}| z_{\rm phot})$ associated with the \dense\ sample and evaluate the correlation functions at the mean redshift of the sample, assuming the fiducial bias and IA amplitude reported in Table~\ref{tab:all_lum_fit_pars}. The ratio is almost constant in $r_{\rm p}$, thus we plot it for fixed $r_{\rm p}=10 \mpch$. We also note that the lensing signal has negligible impact for negative $\Pi$ because the source is in front of the lens in that case.

In principle, if one had perfect knowledge of the galaxy-galaxy lensing contribution, extending the integration up to very large line-of-sight separations would allow us recover the full IA signal from the measurements, without discarding any information. In practice, even though we fully model the galaxy-galaxy lensing contribution, we are limited by the accuracy of the lensing modelling we rely on, and thus it is safer to truncate the integral to values of $\Pi$ that are not severely affected by it.  

We use Fig.~\ref{fig:gG_gI_ratio} to choose the fiducial $\Pi_\mathrm{max}$ that enters in Eq.(~\ref{eq:w_estimator}): although the specific values of the ratio depend on the input parameters ($b_{\rm g}$, $A_\mathrm{IA}$), it provides a realistic estimate of the amount of contamination for our LRG samples. We chose as our fiducial setup a conservative value of $\Pi_{\rm max}= 120 \mpch$, which ensures that the mean contamination is below $\sim 20 \%$ of the signal.

\section{Contamination from magnification } \label{A:magnification_contamination}

The changes in the galaxy number counts determined by lensing magnification arise as a result of two competing effects: on one hand, the lensing locally stretches the sky, diluting the observed number density; on the other hand, it enlarges the apparent sizes of the
galaxies without modifying the surface brightness: at the faint end, this allows the detection of galaxies that are intrinsically fainter than the magnitude limit, enhancing the observed number density.

The theory of magnification for flux-limited surveys is well established and allows us to relate the changes in the number density to the differential galaxy count $n(m)$ over a given band magnitude range from $m$ to $m + \dd m$ \citep{Bartelmann&Schneider2001, JoachimiBridle2010J}:
\begin{equation} \label{eq:alpha_obs}
\alpha(m) = 2.5 \frac{\dd \log[n(m)]}{\dd m} \ .
\end{equation}
The case of a non-flux-limited sample, such as our LRG sample, is more complicated and we lack a proper theoretical framework for the interpretation of $\alpha$. Here, we follow \citet{von-Wietersheim-Kramsta2021} and calibrate $\alpha$ using dedicated mocks, which we present in Appendix~\ref{A:mocks}. We remind the reader that our samples are selected by imposing a luminosity threshold, which implies a redshift-dependent magnitude selection.

The calibration works as follows: the mocks provide the reference relation between the convergence $\kappa$ and the slope $\alpha$, which we can measure as the difference in the number density of a 'magnified' sample and a 'non-magnified' one,
\begin{equation}\label{eq:magnification_number_counts}
    \frac{n(<m) - n_0(<m)}{n_0(<m)} \approx 2(\alpha -1) \kappa \ .
\end{equation}

Here, $n(<m)$ is the local number density of magnified sources with magnitudes below $m$, while $n_0(<m)$ is the underlying true number density without the enhancement due to the flux magnification and the simultaneous lensing dilution.

We used our mocks to measure $\alpha$ in Eq. (\ref{eq:magnification_number_counts}), obtained as the mean value of $\kappa$ on sufficiently small patches of the sky. To partition the sky we use the public available python module \textsc{Healpy}\footnote{\url{https://healpy.readthedocs.io/en/latest/}} \citep{Zonca2019Healpy}, based on the HEALPix pixellization of the sphere\footnote{\url{http://healpix.sourceforge.net}} \citep{Gorski2005Healpix}. We use this value of $\alpha$ to calibrate the magnitude range over which the observable $\alpha$ in equation \ref{eq:alpha_obs} best agrees with the \textit{true} one obtained from equation \ref{eq:magnification_number_counts}. If the mocks reproduce the data selection function to good accuracy, this provides the optimal magnitude range to use to measure $\alpha$ via observable quantities (Eq. (\ref{eq:alpha_obs})) in the data. 

To evaluate equation ~\ref{eq:alpha_obs} we use the $r-$band magnitude and we ensured that the magnitude distribution of the mocks and the data agree to high accuracy. We find that, when applied to the data, the method results in values of $\alpha$ that depend somewhat on the binning scheme employed along the redshift baseline.  While the values of $\alpha$ are robust against changing the bins at intermediate and high redshifts, the very low-$z$ bins are poorly constrained by the method. However, at such low redshifts magnification is negligible, and our samples contain only 
a few galaxies, so it is reasonable to expect the same value of $\alpha$ to hold for the entire sample. Moreover the LRG selection ensures a constant comoving number density, which reduces the sensitivity to magnification even further.

We find $\alpha \sim 1.5$ for both our \dense\ and \lum\ sample. In Appendix~\ref{A:systematic_tests} we show that the effect of including magnification is subdominant in our analysis.

\section{Systematic tests and significance of the detection} \label{A:systematic_tests}

\begin{table*}
	\caption{Reduced $\chi^2$ statistics to assess the significance of our signals $\wgp$ and $w_\mathrm{g \times}$ against the null hypothesis.}
	\label{tab:chi2_null_hypothesis}
		\centering
	\begin{tabular}{lcccc}
		\hline
		\hline
		Sample & Shapes & Signal & $ \chi^2_\mathrm{\nu, null}$ & $p-$value \\
		\hline
        \dense & \textsc{DEIMOS} & $\wgp $ & 8.01 (7.56) & $4.88\times10^{-13}$ ($4.35 \times 10^{-6}$)\\
        & & $\wgx$ & 0.59 (0.36) & 0.83 (0.83) \\
        &  \lensfit & $\wgp$ & 3.52 (4.99) &  0.0001 (0.0005)\\
       & & $\wgx$ & 0.66 (0.64) & 0.76 (0.63) \\
        \hline
        \lum & \textsc{DEIMOS} & $\wgp$ & 9.46 (5.85) & $6.66\times10^{-16}$ (0.0001) \\
        & & $\wgx$ & 0.37 (0.33) & 0.96 (0.85)\\
        &  \lensfit & $\wgp$ & 2.48 (1.49) & 0.006 (0.20) \\
        & & $\wgx$ & 0.40 (0.40) & 0.95 (0.81)\\
		\hline
	\end{tabular}
	\small
	\tablefoot{A detection of $w_\mathrm{g \times}$ would hint at the presence of unaccounted systematics in the measurements. The numbers in brackets refer to the signal for $r_{\rm p} > 6 \ \mathrm{Mpc}/h$.}
\end{table*}

To ensure the robustness of our analysis, we perform a number of tests 
for residual systematics. We present the results of these in this Appendix. Many of these are commonly used to test weak gravitational lensing signals.

In one of the most basic tests the galaxy shapes are rotated by $45\deg$ and the correlation between $\epsilon_{\times}$ and galaxy position, $w_\mathrm{g\times}$ is measured. This correlation is expected to vanish, and any detection of a non-vanishing signal is therefore an indication of residual systematics. Table \ref{tab:chi2_null_hypothesis} reports the reduced $\chi^2_{\nu, \mathrm{null}}$, which we used to assess the significance of the signal against the null hypothesis for both $\wgp$ and $\wgx$. We choose a significance level of $5\%$: for $p-$values below 0.05 we discard the null hypothesis. We can see that all of our $\wgx$ measurements have a $p-$value above 0.05 and thus support the null hypothesis. In contrast, we observe a significant detection for all of our $\wgp$ measurements, for both \textsc{DEIMOS} and \lensfit\ shapes.  

As a further look into possible systematics in the data, we measure the signal for a very large value of the line-of-sight truncation, $\Pi_\mathrm{max}=1000 \mpch$, using our \dense\ sample. Extending the value of $\Pi_\mathrm{max}$ to very large separations introduces uncorrelated pairs into the estimator, and thus we expect the IA signal to vanish, while the galaxy-galaxy lensing can potentially arise. We find a signal consistent with a null detection, with $\chi^2_{\nu,\mathrm{null}} = 0.35$ and $p-$value of 0.96. 

\begin{table*}
	\caption{Tests of the modelling setup.}
	\label{tab:setup_tests}
	\centering
	\begin{tabular}{lccr}
		\hline
		\hline
Sample & $b_\mathrm{g}$ & $A_\mathrm{IA}$ & $\chi^2_\mathrm{red}$\\
\hline
		\textsc{DEIMOS} & & & \\
		\hline 
    	\dense\ (120, baseline) & $1.59^{+0.04}_{-0.04}$ & $3.69^{+0.66}_{-0.65}$ & 0.78 \\
        \dense\ (90) & $1.60^{+0.04}_{-0.04}$ & $3.99^{+0.73}_{-0.72}$ & 0.67 \\
        \dense\ (180) & $1.58^{+0.05}_{-0.05}$ & $3.50^{+0.69}_{-0.70}$ & 0.71 \\
        \dense\ (120, w/o magnification) & $1.59^{+0.04}_{-0.04}$ & $3.67^{+0.66}_{-0.64}$ & 0.78 \\
        \dense\ (120, w/o lensing and magnification) & $1.59^{+0.04}_{-0.04}$ & $3.47^{+0.67}_{-0.66}$ & 0.78 \\
        \lum\ (120, baseline) & $2.06^{+0.04}_{-0.04}$ & $4.03^{+0.81}_{-0.79}$ & 1.19 \\
        \lum\ (120, w/o magnification) & $2.06^{+0.04}_{-0.04}$ & $4.01^{+0.82}_{-0.81}$ & 1.19 \\ 
        \lum\ (120, w/o lensing and magnification)  & $2.06^{+0.04}_{-0.04}$ & $3.84^{+0.80}_{-0.80}$ & 1.19 \\
    \hline
	\end{tabular}
	\small
	\tablefoot{The value of $\Pi_\mathrm{max}$ adopted for the measurement is reported in brackets. The same value is assumed in the model. The tests are always performed using the \textsc{DEIMOS} shape catalogue.}
\end{table*}

We also investigated the impact of specific choices for the setup of our modelling, with a particular focus on how our results depend on the value of the $\Pi_\mathrm{max}$ adopted in the analysis. To do so, we repeat our analysis of the \dense\ sample using two different values of $\Pi_\mathrm{max}$: 90 and 180 $\mpch$. Table \ref{tab:setup_tests} reports our results. We find compatible results that also agree with our fiducial value of $\Pi_\mathrm{max}=120 \mpch$. 

In Table \ref{tab:setup_tests} we also report the results when we include magnification in the modelling for the \dense\ sample, or ignore it for the \lum\ sample.  The resulting parameter estimates agree with the baseline results (also see Sect.~\ref{sec:results}), suggesting magnification is small in our data, as expected from theory \citep[][]{Unruh2020}.

\section{IA dependence on the shape measurement method} \label{A:DEIMOS_lensfit_comparison}

\citet{Singh2016} compared the IA signal measured with different shape methods and found that the signal depends on the specific algorithm employed. \cite{Georgiou2019} explored this further, and used \textsc{DEIMOS} to show that the IA signal depends on the  width of the weight function. Since different methods use different weight functions, the difference in the IA detection can be linked to the parts of the galaxies they probe.

In this Appendix, we therefore explore how the IA signal depends on the shape measurement methods used in our analysis. To ensure this is done consistently, we only select galaxies that belong to both our \textsc{DEIMOS} and \lensfit\ catalogues, irrespective whether they are part of the \lum\ or \dense\ sample. We identify 173\,499 galaxies in common between the two shape catalogues.

\begin{figure}
\centering
\includegraphics[width=\columnwidth]{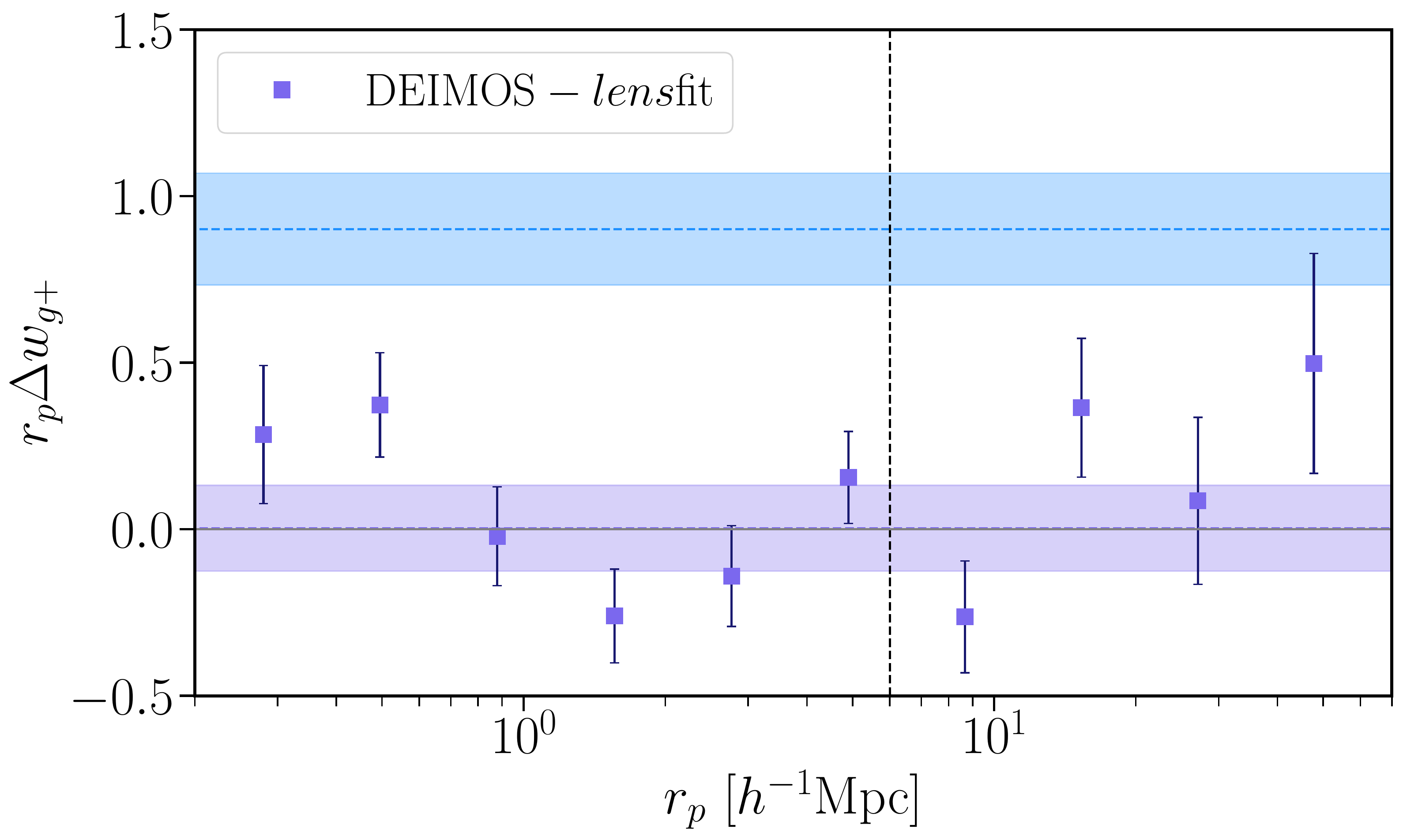}
\caption{Difference in the $\wgp$ measurements as measured by \textsc{DEIMOS} and \lensfit. The indigo dashed line shows the best-fit amplitude of the difference, here parametrised as $~A/r_p$. Similarly, the light blue dashed line illustrates the best-fit amplitude for the \textsc{DEIMOS} sample, both performed for $r_p>6 \mathrm{Mpc}/h$. The shaded areas delimit the $1\sigma$ contour of the fit. }
\label{fig:wgp_lensfit_vs_DEIMOS}
\end{figure}

We measure $\wgp$ for this sub-sample for both shape catalogues,
and show the difference in the signal, $\Delta \wgp = w_\mathrm{g+, DEIMOS}-w_{\mathrm{g}+, lens\mathrm{fit}}$ (indigo squared markers) in Fig.~\ref{fig:wgp_lensfit_vs_DEIMOS}. The error bars are computed via bootstrap; we are only interested in the shape noise contribution: we are measuring the difference of signals obtained using the same sample of galaxies and thus the sample variance should vanish. We generate 215 re-samplings with replacement of our input galaxies and provide the same input catalogue to both our \textsc{DEIMOS} and \lensfit\ measurement of $\wgp$. The error bars are then computed as the standard deviation of the difference in the measured signal for this ensemble. 

To quantify the amplitude of the signal to the potential differences in measurement method, we fit both $w_\mathrm{g+, DEIMOS}$ and $\Delta \wgp$ with a curve of the form $f(r_p) = A/r_p$, for $r_p>6 \mpch$. The best-fit amplitudes are, respectively, $0.90 \pm 0.17$ and $0.003 \pm 0.13$, which means that we detect a signal that is more than six sigma above the uncertainty due to the choice in the shape measurement algorithm adopted.

\section{Mock catalogues} \label{A:mocks}
\begin{figure*}
\centering
\includegraphics[width=\textwidth]{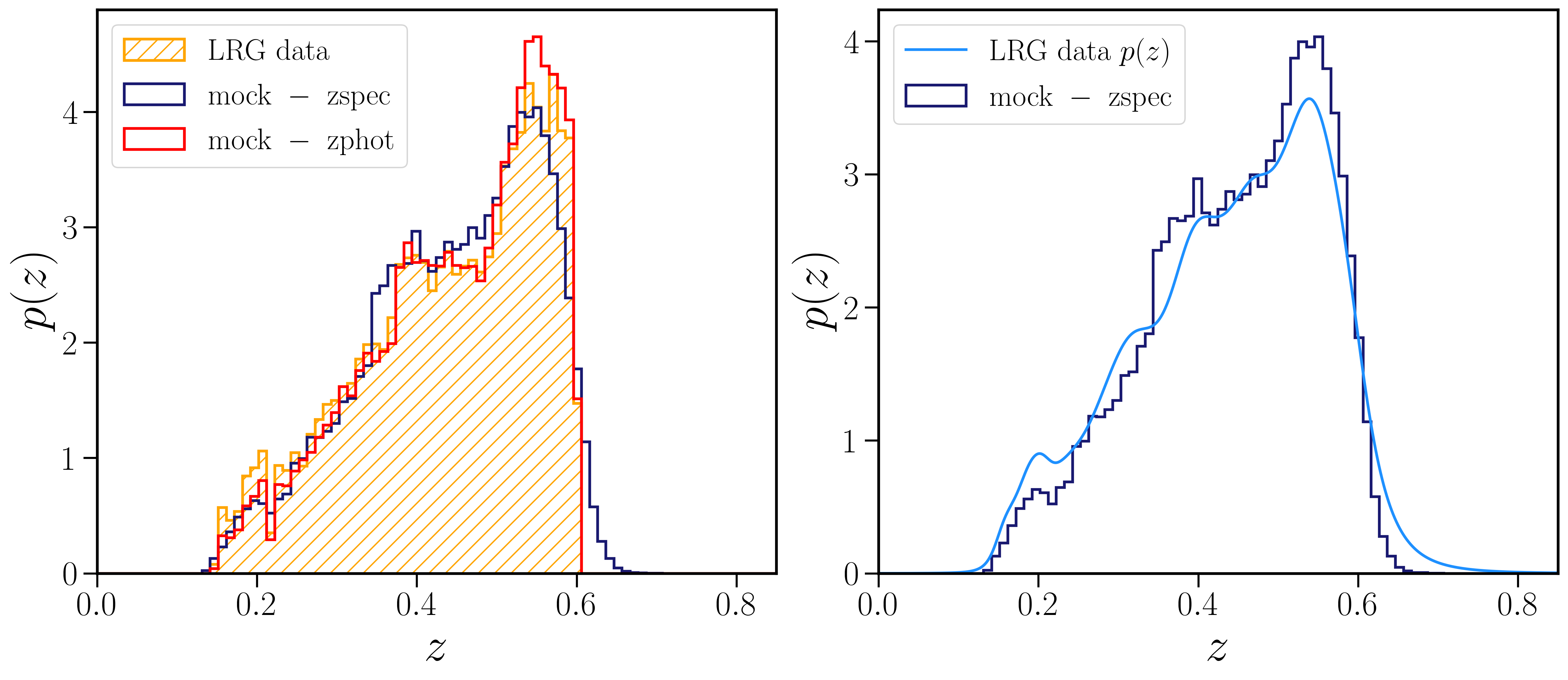}
\caption{Comparison between the redshift distributions of the data and those reproduced by the mocks.  \textit{Left:} The photometric redshift distribution of our data is shown as an orange hatched histogram, while the solid red line shows the distribution of the photometric redshifts of the mock, obtained from the true ('spectroscopic') redshifts (blue solid line) as detailed in the text. \textit{Right:} Comparison of the mock spectroscopic redshift distribution (solid blue line) and the estimated spectroscopic distribution of our data (light blue line).}
\label{fig:mocks_z_distributions}
\end{figure*}

\begin{table}
	\caption{Parameters of the Student's $t$-distributions that best-fit the residuals $(z^\mathrm{phot} - z^\mathrm{spec})/\sigma_z$ of our samples.}
	\label{tab:student_t_pars}
	\begin{tabular}{lccr} 
		\hline
		\hline
		Sample & $\nu$ & $\mu$ & $s$\\
		\hline
        \dense & 3.79 & 0.06 & 0.90 \\
        \lum & 3.99 & $-5.43 \times 10^{-6}$ & 0.86 \\
		\hline
	\end{tabular}
\end{table}

To investigate the impact of magnification bias on the interpretation of our measurements, we generate two mock catalogues that resemble our LRG samples. Our simulated catalogues are obtained from the KiDS photometric mock catalogue presented in \citet[][]{vandenBusch2020}, which is based on the MICEv2 simulation\footnote{\url{http://maia.ice.cat/mice/}} \citep{FosalbaMICEI, CrocceMICEII, FosalbaMICEIII, Carretero2017CosmoHub, Hoffmann2015MICE} and is specifically designed to reproduce the KiDS photometry. We did not run the LRG selection algorithm on the mock, but rather used their observed location in the redshift-colour space ($u-g$, $g-r$, $r-i$, $i-z$) to select them in the mock.

We first apply a broad colour selection using the MICE \texttt{z$\_$cgal} 'spectroscopic' redshift. After assigning the photo-$z$ to our mocks, we repeat the selection replacing the spectroscopic redshift with the photometric one. The photometric redshifts are designed to reproduce the distributions reported in
Sect.~3.3 of \citet{Vakili2020}. To do so, we draw a random value from a Student's $t-$distribution centred on $z_\mathrm{spec} - \mu \sigma_z$ and with the scale parameter equal to $s \sigma_z$, with $\mu$, $\nu$ and $s$ the Student's $t-$parameters fitted to the full distribution (of the real data). We remind the reader that $\nu$ defines the peakiness of the distribution, $\mu$ its mean and $s$ sets the width. In the limit of the Student's $t-$distribution approaching a Gaussian ($\nu \to \infty$), $s$ can be interpreted as the standard deviation of the distribution.

We note that our samples differ from \citet[][]{Vakili2020}, since we exclude the galaxies that overlap with the \lum\ sample from the \dense\ sample.  We therefore recompute the parameters of the Student's $t-$distributions specifically for our samples and report these in 
Table~\ref{tab:student_t_pars}. Some care has to be taken when assigning $\sigma_z$ to the mocks. The per-galaxy $\sigma_z$ of the LRG samples correlates with the magnitude of the galaxy. 
We therefore identify the closest real galaxy in the $(z,m_r)$ space to each galaxy in the mock, and assign it the corresponding $\sigma_z$. We repeat the process for one iteration, replacing the 'spectroscopic' redshift with the preliminary estimate of the photometric one. We note that this procedure results in multiple assignments of the same $\sigma_z$ to the mock galaxies, but this is not a concern as we do not require it to be unique. 

Since we require a high fidelity reproduction of the line-of-sight distribution of our galaxies, we divide our samples and their corresponding mock catalogues in thin redshift slices and match the galaxy number density per slice. At this step, we do not require a perfect match. In this way, we still have enough galaxies to apply the same $m^{\rm pivot}_{r}(z)$ cut as for our real data. We repeat these steps iteratively until the number densities are matched between the samples. We tested that the final $p(z^\mathrm{spec}|z^\mathrm{phot})$ of our mocks are in good agreement with the data $p(z^\mathrm{spec}|z^\mathrm{phot})$ (see Fig.~\ref{fig:mocks_z_distributions}) and that the resulting clustering signal at large scales reproduces the one in our data. 

We generate two sets of mock catalogues: a magnified one and one without magnification. We use these for the calibration of $\alpha$ as discussed in Appendix \ref{A:magnification_contamination}.


\end{document}